\documentclass[fleqn,usenatbib]{mnras}

\usepackage{newtxtext,newtxmath}

\usepackage[T1]{fontenc}

\DeclareRobustCommand{\VAN}[3]{#2}
\let\VANthebibliography\thebibliography
\def\thebibliography{\DeclareRobustCommand{\VAN}[3]{##3}\VANthebibliography}

\usepackage{textcomp,graphicx,amsmath,setspace,multirow,psfrag,color,orcidlink}

\title[Late-time observations of SN 2021qvv]{No plateau observed in late-time near-infrared observations of the underluminous Type Ia supernova 2021qvv}
\author[O. Graur et al.]{
O.~Graur\orcidlink{0000-0002-4391-6137},$^{1,2}$\thanks{E-mail: \href{mailto:or.graur@port.ac.uk}{or.graur@port.ac.uk}}
E.~Padilla Gonzalez\orcidlink{0000-0003-0209-9246},$^{3,4}$
J.~Burke,$^{3,4}$
M.~Deckers\orcidlink{0000-0001-8857-9843},$^5$
S.~W.~Jha\orcidlink{0000-0001-8738-6011},$^6$
L.~Galbany\orcidlink{0000-0002-1296-6887},$^{7,8}$
\newauthor
E.~Karamehmetoglu\orcidlink{0000-0001-6209-838X},$^9$
M.~D.~Stritzinger\orcidlink{0000-0002-5571-1833},$^9$
K.~Maguire\orcidlink{0000-0002-9770-3508},$^5$
D.~A.~Howell\orcidlink{0000-0003-4253-656X},$^{3,4}$
R. Fisher\orcidlink{0000-0001-8077-7255},$^{10}$
A.~G.~Fullard\orcidlink{0000-0001-7343-1678},$^{11}$
\newauthor
R.~Handberg\orcidlink{0000-0001-8725-4502},$^9$ 
D.~Hiramatsu\orcidlink{0000-0002-1125-9187},$^{12,13}$
G.~Hosseinzadeh\orcidlink{0000-0002-0832-2974},$^{14}$
W.~E.~Kerzendorf\orcidlink{0000-0002-0479-7235},$^{11}$
C.~McCully\orcidlink{0000-0001-5807-7893},$^{3,4}$ 
\newauthor
M.~Newsome\orcidlink{0000-0001-9570-0584},$^{3,4}$
C.~Pellegrino\orcidlink{0000-0002-7472-1279},$^{3,4}$
A.~Rest\orcidlink{0000-0002-4410-5387},$^{15,16}$
A.~G.~Riess\orcidlink{0000-0002-6124-1196},$^{15,16}$
I.~R.~Seitenzahl\orcidlink{0000-0002-5044-2988},$^{17}$
M.~M.~Shara\orcidlink{0000-0003-0155-2539},$^2$
\newauthor
K.~J.~Shen\orcidlink{0000-0002-9632-6106},$^{18}$
G.~Terreran\orcidlink{0000-0003-0794-5982},$^{3,4}$     
and D.~R.~Zurek$^2$
\\
$^1$Institute of Cosmology and Gravitation, University of Portsmouth, Portsmouth, PO1 3FX, UK\\
$^2$Department of Astrophysics, American Museum of Natural History, Central Park West and 79th Street, New York, NY 10024-5192, USA\\
$^3$Las Cumbres Observatory, 6740 Cortona Dr, Suite 102, Goleta, CA 93117-5575, USA\\
$^4$Department of Physics, University of California, Santa Barbara, CA 93106-9530, USA\\
$^5$School of Physics, Trinity College Dublin, College Green, Dublin 2, Ireland\\
$^6$Department of Physics and Astronomy, Rutgers, the State University of New Jersey, 136 Frelinghuysen Road, Piscataway, NJ 08854-8019, USA\\
$^7$Institute of Space Sciences (ICE, CSIC), Campus UAB, Carrer de Can Magrans, s\/n, E-08193 Barcelona, Spain\\
$^8$Institut d'Estudis Espacials de Catalunya (IEEC), E-08034 Barcelona, Spain\\
$^9$Department of Physics and Astronomy, Aarhus University, Ny Munkegade 120, DK-8000 Aarhus C, Denmark \\
$^{10}$Department of Physics, University of Massachusetts Dartmouth, 285 Old Westport Road, North Dartmouth, MA 02740, USA \\
$^{11}$Department of Physics and Astronomy, Michigan State University, East Lansing, MI 48824, USA \\
$^{12}$Center for Astrophysics | Harvard \& Smithsonian, 60 Garden Street, Cambridge, MA 02138-1516, USA \\
$^{13}$The NSF AI Institute for Artificial Intelligence and Fundamental Interactions, USA \\
$^{14}$Steward Observatory, University of Arizona, 933 North Cherry Avenue, Tucson, AZ 85721-0065, USA \\
$^{15}$Space Telescope Science Institute, Baltimore, MD 21218, USA \\
$^{16}$Department of Physics and Astronomy, The Johns Hopkins University, Baltimore, MD 21218, USA \\
$^{17}$School of Science, University of New South Wales Canberra, The Australian Defence Force Academy, Canberra, ACT 2600, Australia \\
$^{18}$Department of Astronomy and Theoretical Astrophysics Center, University of California, Berkeley, CA 94720, USA
}

\date{Accepted 2023 September 26. Received 2023 September 14; in original form 2023 June 22}

\pubyear{2023}

\begin{document}
\label{firstpage}
\pagerange{\pageref{firstpage}--\pageref{lastpage}}
\maketitle



\begin{abstract}
\noindent Near-infrared (NIR) observations of normal Type Ia supernovae (SNe~Ia) obtained between 150 to 500 d past maximum light reveal the existence of an extended plateau. Here, we present observations of the underluminous, 1991bg-like SN 2021qvv. Early, ground-based optical and NIR observations show that SN 2021qvv is similar to SN 2006mr, making it one of the dimmest, fastest-evolving 1991bg-like SNe to date. Late-time (170--250 d) \textit{Hubble Space Telescope} observations of SN 2021qvv reveal no sign of a plateau. An extrapolation of these observations backwards to earlier-phase NIR observations of SN 2006mr suggests the complete absence of a NIR plateau, at least out to 250 d. This absence may be due to a higher ionization state of the ejecta, as predicted by certain sub-Chandrasekhar-mass detonation models, or to the lower temperatures of the ejecta of 1991bg-like SNe, relative to normal SNe Ia, which might preclude their becoming fluorescent and shifting ultraviolet light into the NIR. This suggestion can be tested by acquiring NIR imaging of a sample of 1991bg-like SNe that covers the entire range from slowly-evolving to fast-evolving events ($0.2 \lesssim s_\mathrm{BV} \lesssim 0.6$). A detection of the NIR plateau in slower-evolving, hotter 1991bg-like SNe would provide further evidence that these SNe exist along a continuum with normal SNe Ia. Theoretical progenitor and explosion scenarios would then have to match the observed properties of both SN Ia subtypes.

\end{abstract}



\begin{figure*}
 \includegraphics[width=0.93\textwidth]{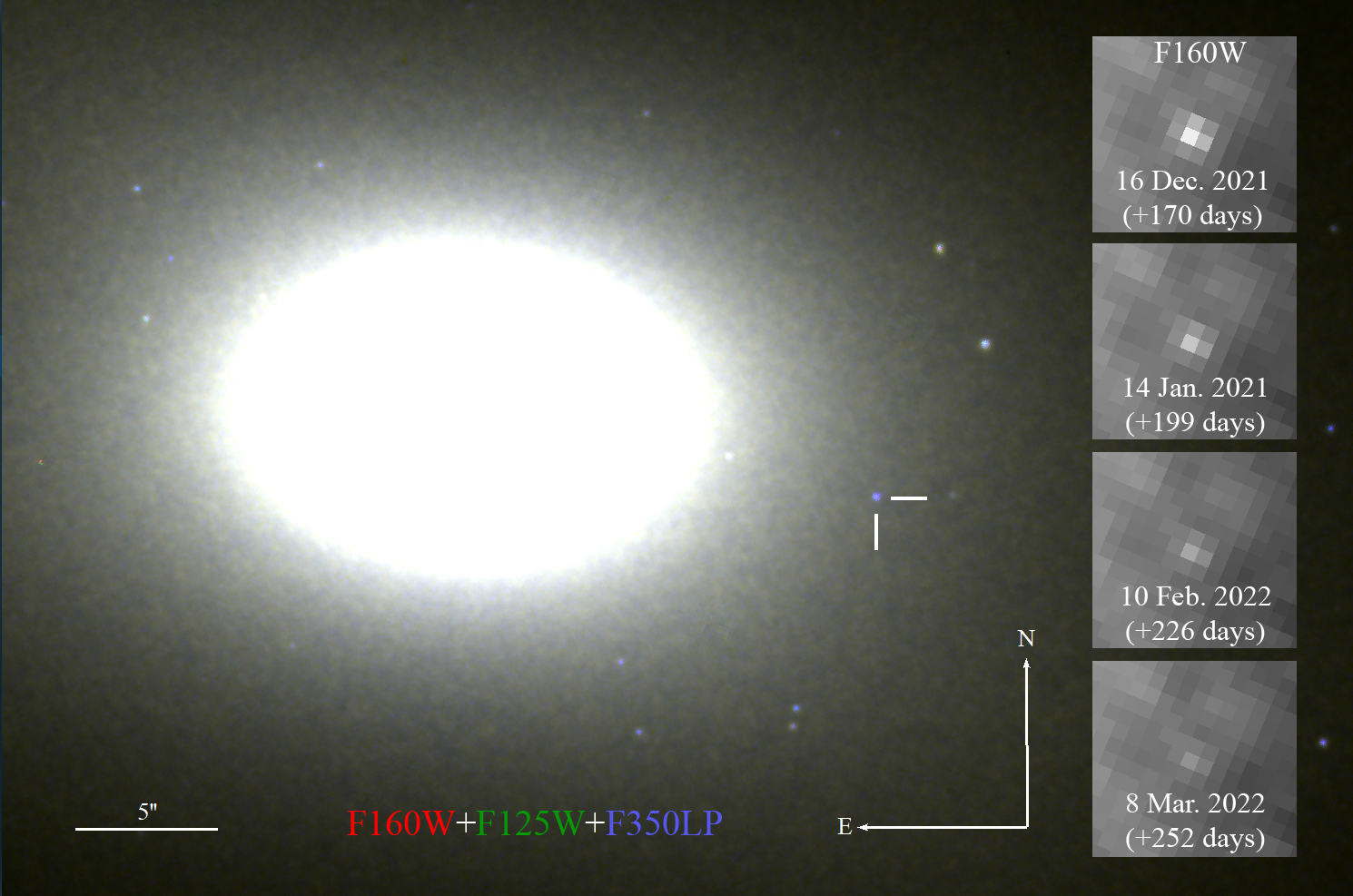}
 \caption{Multicolored \textit{HST} WFC3 image of SN 2021qvv (in the white reticle) and its lenticular host galaxy, NGC 4442. The cutouts show the SN declining in brightness in the WFC3 \textit{F160W} band over a time span of four months, $\sim 170$--$250$ d post maximum light.}
 \label{fig:gal}
\end{figure*}

\section{Introduction}
\label{sec:intro}

Though Type Ia supernovae (SNe Ia) are widely used as standardisable candles in cosmology, we still do not know what type of star (or star system) explodes as this type of SN and how the explosion occurs (see reviews by \citealt{2012NewAR..56..122W,2014ARA&A..52..107M,2019NatAs...3..706J,2019NewAR..8701535S,2023RAA....23h2001L}; for a general introduction to SNe, see \citealt{2022supe.book.....G}). This effort is compounded by the heterogeneity of SNe Ia. So-called `normal' SNe Ia share the Universe with overluminous 1991T-like SNe \citep{1992ApJ...384L..15F}, underluminous 1991bg-like SNe \citep{1992AJ....104.1543F}, transitional iPTF13ebh-like SNe \citep{2015AA...578A...9H}, peculiar SNe Iax \citep{2006AJ....132..189J,2013ApJ...767...57F}, and other SN Ia subtypes with their own peculiarities. 

Late-time observations of SNe Ia taken $\sim150$--$2000$ d past maximum $B$-band light have come forth as a probe of the progenitor and explosion mechanism of these SNe. Ground-based, optical observations out to $\approx 640$ d have shown a continuous decline \citep{2004A&A...428..555S,2006AJ....132.2024L,2009A&A...505..265L}, as opposed to the steep drop predicted by the so-called `infrared catastrophe' \citep{1980PhDT.........1A}. Optical observations taken with the \textit{Hubble Space Telescope} (\textit{HST}) starting at $\approx 800$ d have further revealed a flattening of the SN Ia light curve that cannot be attributed to contaminating light echoes or contribution from a surviving companion (\citealt{2016ApJ...819...31G,2017ApJ...841...48S,2017ApJ...834..180S,2018ApJ...852...89Y,2019ApJ...882...30L,2022MNRAS.517.4119T}; though see \citealt{2018ApJ...854...55Y} regarding the use of imaging polarimetry to discover light echoes unresolvable in regular imaging).   

\citet{2018ApJ...859...79G,2018ApJ...866...10G} and \citet{2019ApJ...870...14G} have further claimed that the shape of the late-time light curves may be correlated with the peak luminosities of the SNe (though see \citealt{2018ApJ...857...88J}). This is reminiscent of the peak width-luminosity relation used to standardize SNe Ia for use as standard candles (see review by \citealt{2017hsn..book.2543P}). If real, a second, late-time width-luminosity correlation could be used to further standardise SNe Ia and lower the systematic uncertainties associated with their use as cosmology probes. However, a larger sample of SNe is required to test the significance of this correlation.

Near-infrared (NIR) observations of SNe Ia between 150 and 500 d have revealed a year-long plateau in the $J$- and $H$-band light curves \citep{2004A&A...428..555S,2004A&A...426..547S,2006AJ....132.2024L,2007A&A...470L...1S,2020NatAs...4..188G,2023MNRAS.tmp..775D}. This plateau is thought to be the result of ultraviolet (UV) light being shifted into the optical and NIR \citep{2004A&A...428..555S,2015ApJ...814L...2F,2020NatAs...4..188G}. To date, only normal SNe Ia have been observed long enough to detect this plateau, and it is unclear whether other SN Ia subtypes exhibit a similar NIR plateau or not. In this paper, we present the first late-time observations of a 1991bg-like SN Ia, SN 2021qvv, which show---at least out to 250 d---no evidence of a NIR plateau. 

In Section~\ref{sec:obs}, we describe our observations of SN 2021qvv, which consist of ground-based imaging and spectroscopy around peak, along with \textit{HST} NIR imaging at 170--250 d past maximum light. We only have one epoch of NIR observations around peak light, so we combine literature NIR data of SN 2006mr with our late-time \textit{HST} data. This is not a simple operation, as 1991bg-like SNe are a heterogeneous family. Hence, in Section~\ref{sec:analysis}, we first demonstrate that SN 2021qvv is a bona fide 1991bg-like SN, before showing that SN 2006mr is its closest relative. With SN 2021qvv stitched on to SN 2006mr, we find no indication of a late-time NIR plateau out to 250 d. 

In Section~\ref{sec:discuss}, we pin the absence of a NIR plateau on the temperature of the ejecta of 1991bg-like SNe, which is known to be appreciably lower than in normal SNe Ia. If this is the case, then the lack of a plateau in SN 2021qvv does not necessarily mean that \emph{all} 1991bg-like SNe lack a NIR plateau. SNe 2006mr and 2021qvv are two of the fastest-evolving and dimmest 1991bg-like SNe observed to date, and 1991bg-like SNe exist on a continuum, with slower-evolving, more luminous members of this family exhibiting hotter ejecta and NIR features progressively more similar to those of normal SNe Ia. Hence, it is plausible that slower, hotter 1991bg-like SNe will exhibit late-time NIR plateaus. Finally, we summarise our conclusions in Section~\ref{sec:conclude}, where we encourage further late-time observations of 1991bg-like SNe.


\section{Observations}
\label{sec:obs}

SN 2021qvv was discovered by the Asteroid Terrestrial-impact Last Alert System (ATLAS; \citealt{2018PASP..130f4505T}) on 2021 June 23 \citep{2021TNSTR2173....1T}. It is located $14\farcs3$ west and $3\farcs3$ south of the center of NGC 4442, a nearby lenticular galaxy. The SN and its host galaxy are shown in Figure~\ref{fig:gal}. We obtained optical and NIR imaging of SN 2021qvv with the Las Cumbres Observatory, the 3.5-m telescope at the Calar Alto Observatory, and \textit{HST} (Section~\ref{subsec:imaging}), as well as three optical spectra around peak (Section~\ref{subsec:spec}).

\begin{table*}
 \caption{Las Cumbres Observatory observation log of SN 2021qvv}\label{table:LCO}
 \begin{tabular}{lcccccccc}
  \hline
  \hline
  Date         & MJD     & Phase  & $U$           & $B$           & $V$           & $g$          & $r$            & $i$ \\
               & (d)     & (d)    & (Vega mag)    & (Vega mag)    & (Vega mag)    & (AB mag)     & (AB mag)       & (AB mag) \\
  \hline
  24 Jun. 2021 & 59390.3 & $-5.0$ & $\cdots$      & 15.337(0.021) & 14.969(0.025) & $\cdots$      & 15.197(0.015) & 15.307(0.019) \\
  26 Jun. 2021 & 59391.9 & $-3.4$ & 14.470(0.058) & 14.787(0.010) & 14.512(0.014) & 14.502(0.004) & 14.518(0.004) & $\cdots$      \\
  27 Jun. 2021 & 59393.6 & $-1.7$ & $\cdots$      & $\cdots$      & 14.249(0.020) & $\cdots$      & 14.316(0.006) & $\cdots$      \\
  28 Jun. 2021 & 59394.5 & $-0.8$ & $\cdots$      & 14.549(0.014) & 14.138(0.020) & $\cdots$      & 14.235(0.006) & 14.440(0.007) \\
  29 Jun. 2021 & 59395.5 & $0.2$  & $\cdots$      & 14.512(0.015) & $\cdots$      & $\cdots$      & $\cdots$      & $\cdots$      \\
  30 Jun. 2021 & 59396.5 & $1.2$  & $\cdots$      & $\cdots$      & $\cdots$      & $\cdots$      & 13.892(0.007) & $\cdots$      \\
   1 Jul. 2021 & 59397.5 & $2.2$  & $\cdots$      & 14.592(0.017) & 13.803(0.022) & $\cdots$      & 13.848(0.008) & 14.074(0.013) \\
   2 Jul. 2021 & 59398.5 & $3.2$  & $\cdots$      & $\cdots$      & $\cdots$      & $\cdots$      & $\cdots$      & 13.90(0.11)   \\
   3 Jul. 2021 & 59399.4 & $4.1$  & 15.372(0.059) & 15.019(0.011) & 13.923(0.014) & 14.498(0.004) & 13.768(0.004) & 13.845(0.004) \\
   4 Jul. 2021 & 59400.5 & $5.2$  & 15.657(0.061) & 15.182(0.011) & 14.044(0.014) & 14.654(0.004) & 13.793(0.004) & 13.913(0.005) \\
   6 Jul. 2021 & 59402.3 & $7.0$  & 15.802(0.066) & 15.607(0.012) & 14.240(0.014) & 15.005(0.005) & 13.938(0.004) & 13.981(0.005) \\
   7 Jul. 2021 & 59403.5 & $8.2$  & 16.243(0.064) & 15.783(0.012) & 14.474(0.014) & 15.244(0.005) & 14.135(0.004) & 14.156(0.005) \\
   9 Jul. 2021 & 59405.5 & $10.2$ & 16.387(0.065) & 16.072(0.013) & 14.754(0.014) & 15.551(0.005) & 14.378(0.004) & 14.310(0.005) \\
  11 Jul. 2021 & 59407.2 & $11.9$ & 16.476(0.098) & 16.249(0.016) & 14.885(0.015) & 15.671(0.007) & 14.564(0.005) & 14.442(0.006) \\
  13 Jul. 2021 & 59409.5 & $14.2$ & 16.61(0.11)   & 16.533(0.019) & 15.154(0.015) & $\cdots$      & 14.820(0.006) & 14.720(0.007) \\
  17 Jul. 2021 & 59413.5 & $18.2$ & 16.60(0.10)   & 16.644(0.041) & 15.246(0.018) & $\cdots$      & 15.053(0.010) & 14.945(0.009) \\
  19 Jul. 2021 & 59415.2 & $19.5$ & $\cdots$      & 16.905(0.080) & 15.495(0.054) & $\cdots$      & $\cdots$      & $\cdots$      \\
  20 Jul. 2021 & 59416.5 & $21.2$ & 16.94(0.18)   & 16.710(0.023) & 15.536(0.020) & $\cdots$      & 15.278(0.011) & 15.256(0.012) \\
  21 Jul. 2021 & 59418.5 & $23.2$ & $\cdots$      & 17.021(0.038) & $\cdots$      & $\cdots$      & $\cdots$      & $\cdots$      \\
  23 Jul. 2021 & 59419.5 & $24.2$ & 16.74(0.27)   & 16.950(0.120) & 15.684(0.025) & $\cdots$      & 15.610(0.016) & 15.448(0.015) \\
  26 Jul. 2021 & 59422.2 & $26.9$ & 16.94(0.14)   & 17.122(0.023) & 15.897(0.022) & $\cdots$      & 15.835(0.015) & 15.615(0.017) \\
  30 Jul. 2021 & 59425.5 & $30.2$ & 17.11(0.10)   & $\cdots$      & 16.124(0.022) & $\cdots$      & 16.097(0.017) & 15.881(0.016) \\
  \hline
 \end{tabular}
\end{table*}

\begin{table*}
 \caption{\textit{HST} observation log of SN 2021qvv}\label{table:HST}
 \begin{tabular}{lccccc}
  \hline
  \hline
  Date & MJD    & Phase  & \textit{F350LP} & \textit{F125W} & \textit{F160W} \\
       & (d)    & (d)    & (Vega mag)      & (Vega mag)     & (Vega mag)     \\
  \hline
  16 Dec. 2021  & 59564.8 & 169.5 & 20.381(0.011) & 20.529(0.014) & 19.939(0.011) \\
  14 Jan. 2022  & 59594.0 & 198.7 & 20.989(0.016) & 21.069(0.014) & 20.386(0.017) \\
  10 Feb. 2022  & 59621.0 & 225.7 & 21.658(0.023) & 21.457(0.019) & 20.947(0.026) \\
  08 Mar. 2022  & 59647.0 & 251.7 & 22.110(0.028) & 21.920(0.220) & 21.554(0.045) \\
  \hline
 \end{tabular}
\end{table*}

\subsection{Imaging}
\label{subsec:imaging}

We observed SN 2021qvv between $-5.0$ and $30.2$ d past $B$-band maximum light using Las Cumbres Observatory's network of 1-m telescopes \citep{2013PASP..125.1031B} as part of the Global Supernova Project and the Aarhus-Barcelona cosmic FLOWS project.\footnote{\href{https://flows.phys.au.dk/}{https://flows.phys.au.dk/}} Due to the object's proximity to its host, we performed image subtraction using \texttt{HOTPANTS} \citep{hotpants} and template images taken by Las Cumbres Observatory on 2022 May 26 after the object had faded. Photometry of the SN was then obtained from the difference images using \texttt{lcogtsnpipe} \citep{2016MNRAS.459.3939V}, a PyRAF-based pipeline.

\begin{figure}
 \includegraphics[width=0.475\textwidth]{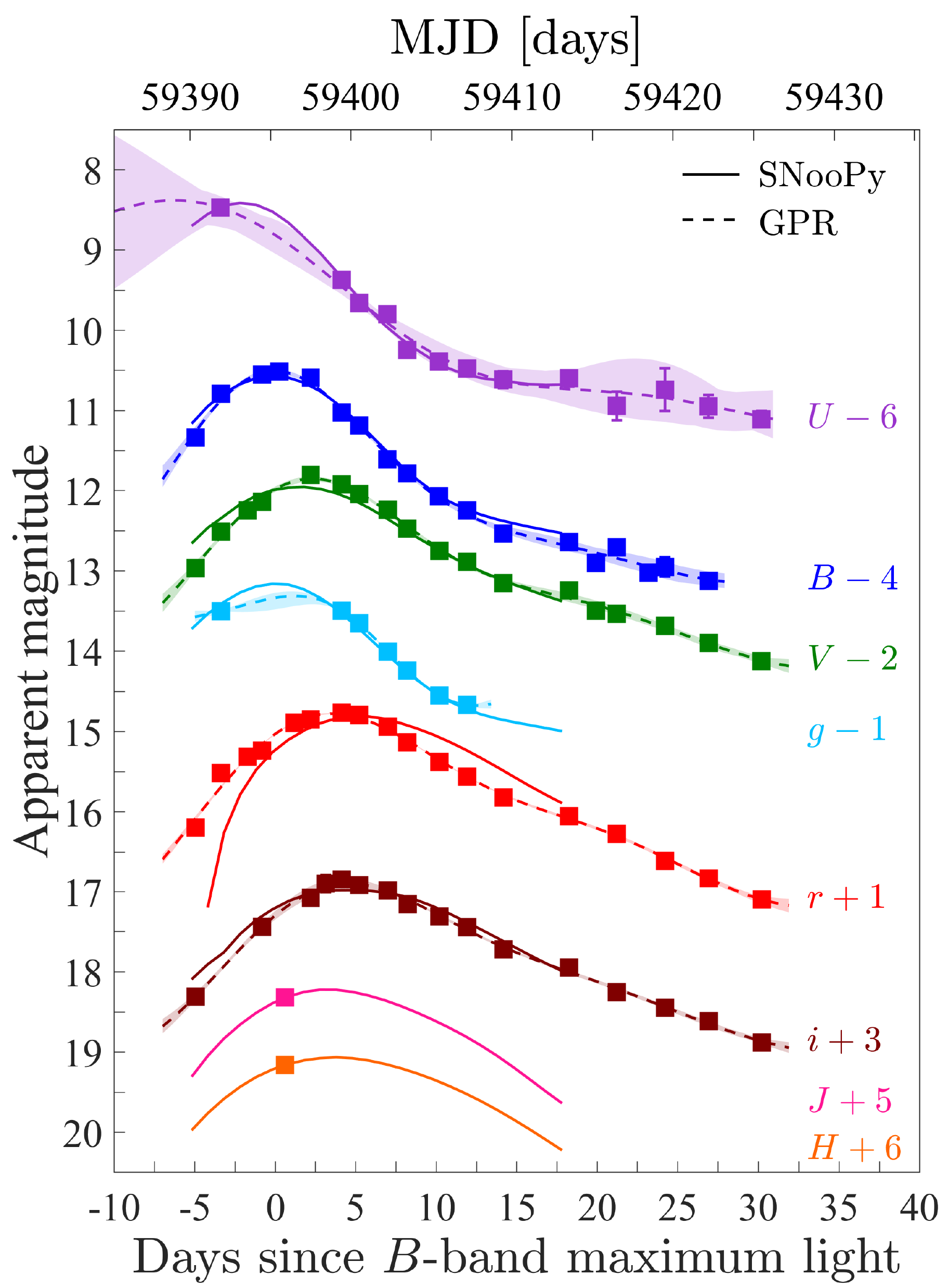}
 \caption{Ground-based optical and NIR light curves of SN 2021qvv obtained with the Las Cumbres Observatory's network of 1-m telescopes and the 3.5-m telescope at the Calar Alto Observatory, respectively. The solid curves are SNooPy fits, while the dashed curves and patches are the GPR fits to the observations and their $1\sigma$ uncertainty regions. The SNooPy $1\sigma$ uncertainty regions are too small to show up in the figure.}
 \label{fig:lc_LCO}
\end{figure}

\textit{UBV} observations are reported in Vega mags. We calculate zeropoints for Landolt filters using same-night, same-telescope observations of Landolt standard fields \citep{stetson}. \textit{gri} observations are reported in AB mags. The zeropoints in these bands were calculated using Sloan catalog stars in the fields of the SN \citep{2017ApJS..233...25A}. The final light curves are presented in Table~\ref{table:LCO} and Figure~\ref{fig:lc_LCO}.

\begin{figure*}
 \begin{tabular}{cc}
  \includegraphics[width=0.475\textwidth]{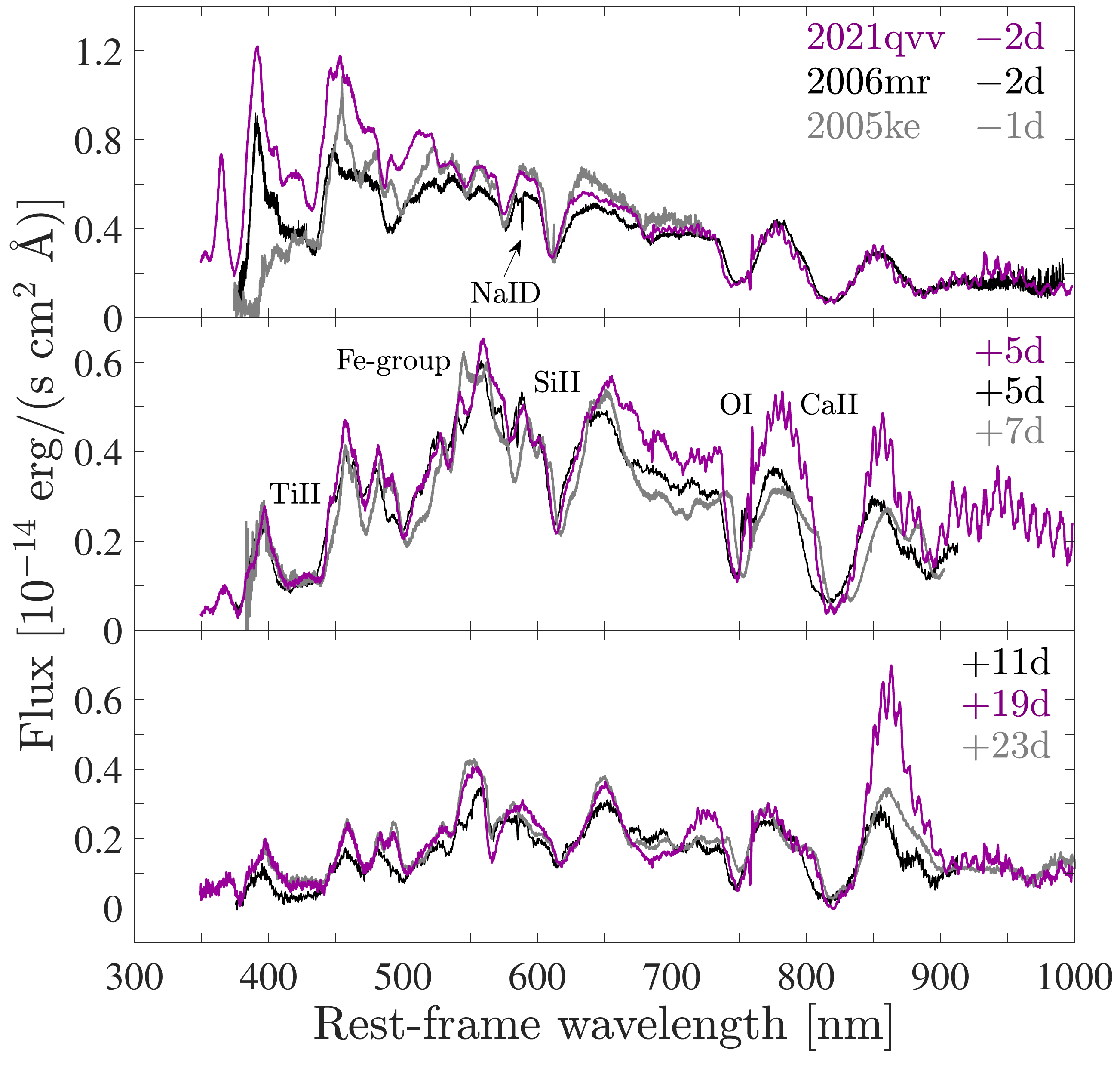} &
  \includegraphics[width=0.475\textwidth]{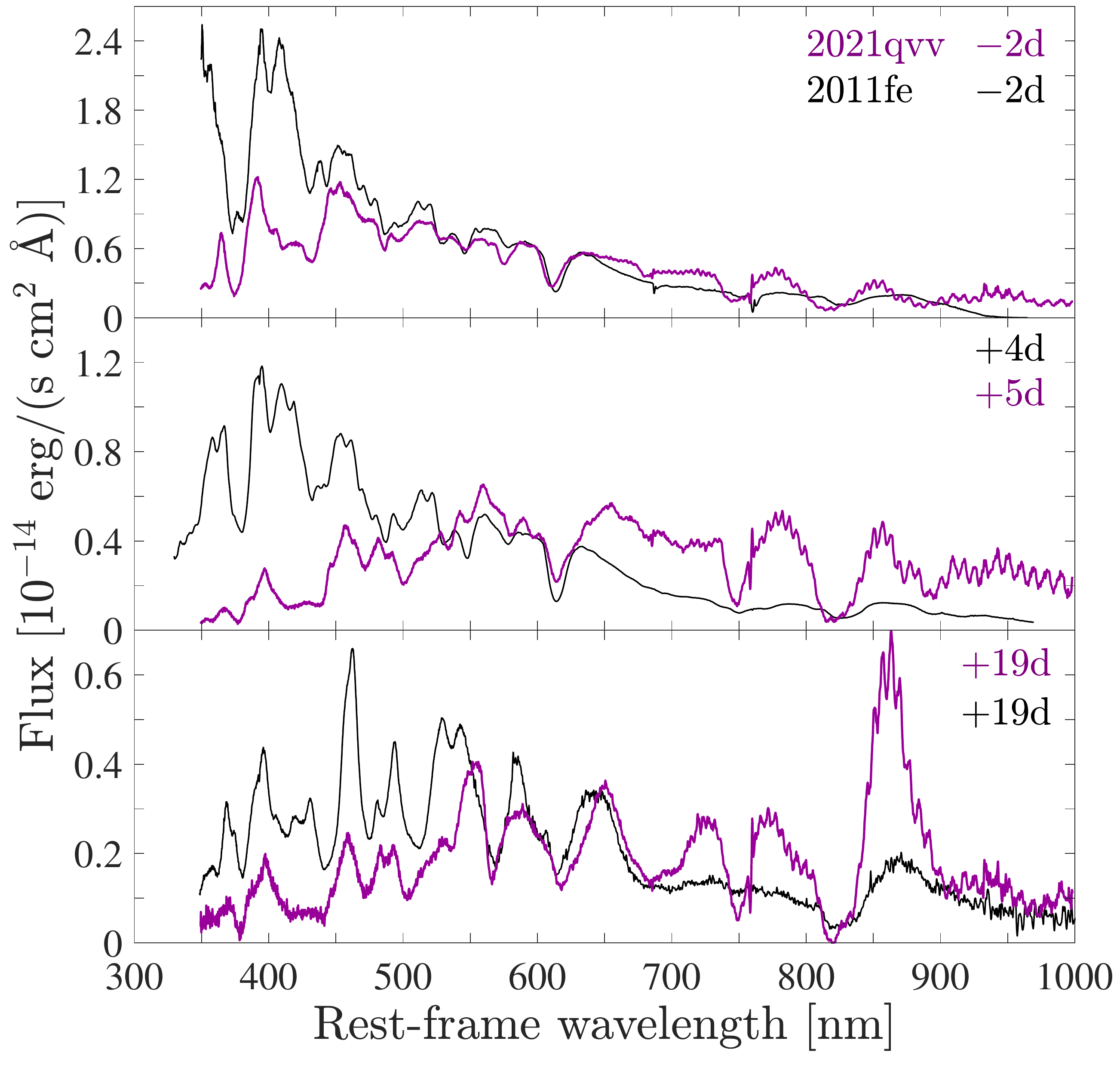} 
 \end{tabular}
 \caption{A spectroscopic time series of SN 2021qvv (purple), compared to the 1991bg-like SNe 2005ke and 2006mr (left) and the normal SN Ia 2011fe (right). The spectra of the comparison SNe have been normalized to the spectra of SN 2021qvv at the 635.5 nm Si~\textsc{ii} absorption feature. The resemblance of SN 2021qvv to SN 2006mr is apparent, as is the marked difference between SNe 2021qvv and 2011fe. Note, especially, the Ti~\textsc{ii} band at 400--440 nm, which is characteristic of 1991bg-like SNe Ia. Also note the narrow Na ID line in the pre-maximum spectrum of SN 2006mr, which is due to host-galaxy reddening. This feature is absent from the spectra of SN 2021qvv, which we attribute to an absence of dust at the location of the SN.}
 \label{fig:spec}
\end{figure*}

A single epoch of imaging in the $J$ and $H$ bands was obtained on 2021 June 30 (MJD 59395.9, $0.6$ d after $B$-band maximum light) with the Omega2000 camera mounted on the 3.5-m telescope at the Calar Alto Observatory. Ten and 15 images, each 60 s long, were obtained in the $J$ and $H$ bands, respectively. The images were dark and flat corrected, sky subtracted, and combined using our own IRAF-based pipeline. Point-spread function (PSF) fitting photometry was then computed using the FLOWS project's automated pipeline,\footnote{\href{https://github.com/SNflows/flows}{https://github.com/SNflows/flows}} and calibrated using stars from the Two Micron All Sky Survey \citep{2006AJ....131.1163S}. We note the SN is at a location with a bright galaxy background and that host-galaxy contribution is not excluded due to the lack of host-galaxy templates. While it is possible to attempt to remove the galaxy background by, e.g., fitting the galaxy's light profile, this was not done here. However, as we show in Section~\ref{subsec:plateau}, we estimate the galaxy background to have a minimal effect on our photometry. The resultant photometry, shown in Figure~\ref{fig:lc_LCO}, is $J=13.32 \pm 0.04$ mag and $H=13.16 \pm 0.06$ mag (Vega).

Late-time images of SN 2021qvv were obtained with the \textit{HST} Wide Field Camera 3 (WFC3) wide-band filters \textit{F350LP} (white light), \textit{F125W} ($J$), and \textit{F160W} ($H$) under \textit{HST} program GO--16884 (PI: O. Graur) on four separate occasions between 2021 December 16 and 2022 March 8. During these visits, the SN was $\sim 170$--250 d past maximum $B$-band light. These observations are presented in Table~\ref{table:HST} and can be downloaded from the Mikulski Archive for Space Telescopes (MAST) at doi:\href{http://dx.doi.org/10.17909/psf9-7d70}{10.17909/psf9-7d70}.

During each visit, we aligned and stacked the dithered \textit{HST} exposures in each filter using the \texttt{TWEAKREG} and \texttt{ASTRODRIZZLE} tasks in the \texttt{DRIZZLEPAC} Python Package \citep{2012drzp.book.....G}.\footnote{\href{http://drizzlepac.stsci.edu/}{http://drizzlepac.stsci.edu/}} This process also removes cosmic rays and bad pixels. The location of SN 2021qvv was verified by comparing our observations to pre-explosion images taken with the \textit{HST} Advanced Camera for Surveys on 2003 September 9 (GO-9401; PI: P. Cote). As expected, the SN does not appear in either the \textit{F475W} ($B$) or \textit{F850LP} ($z$) pre-explosion images. 

Next, we performed PSF-fitting photometry using \texttt{DOLPHOT} \citep{2000PASP..112.1383D,2016ascl.soft08013D},\footnote{\href{http://americano.dolphinsim.com/dolphot/}{http://americano.dolphinsim.com/dolphot/}} which is specifically designed to extract photometry from \textit{HST} images. Because the SN fades over time, we used forced photometry on the centroid of the SN as derived by \texttt{DOLPHOT} in the first, brightest, visit. The resultant photometry, in Vega mags, is presented in Table~\ref{table:HST}.

Analysis of the \texttt{DOLPHOT} photometry revealed the first \textit{F125W} visit to be an outlier. In this visit, \texttt{DOLPHOT} recorded crowding of the SN by a second point source. In the first subexposure, \texttt{DOLPHOT} measured a magnitude of $20.029 \pm 0.009$ mag, with a crowding of $0.8$ mag. In the second subexposure, the measured photometry is $20.296 \pm 0.018$ mag and a crowding of $0.3$ mag. Yet a visual inspection of the subexposures revealed no hint of a second point source at or close to the location of the SN.

None of the subexposures from the other \textit{F125W} or any of the \textit{F160W} visits suffer from crowding. \texttt{DOLPHOT} reports some crowding in all of the \textit{F350LP} visits. However, a visual inspection of the \textit{SN} images reveals no other point source at the location of the SN. Moreover, the crowding reported in \textit{F350LP} is ten times smaller than that reported in the first \textit{F125W} image and has a negligible effect on the final photometry.

To address this issue, we allowed \texttt{DOLPHOT} to derive the centroid of the SN in the second visit, and then used forced photometry on the other visits. The photometry of the second and third visits was consistent with that derived initially, but the photometry of the first and fourth visits was fainter. No crowding was reported, but there was a difference of $\sim 0.6$ mag between the two subexposures of the first visit. We attribute this difference to masking of a single pixel in the second subexposure, and take the value derived from the first subexposure, $20.529 \pm 0.014$ mag, as the \textit{F125W} magnitude of the SN during the first visit. The subexposures of the fourth visit were different by $\sim 0.2$ mag, which we attribute to the faintness of the SN at that phase; we adopt this difference as a more conservative estimate of the photometry uncertainty than the formal value derived by \texttt{DOLPHOT}. As a final check, we repeated the process described here with the third visit used to centre the SN. The results were consistent with those derived before.

\subsection{Spectroscopy}
\label{subsec:spec}

A sequence of three optical spectra were taken with the FLOYDS spectrograph mounted on the Las Cumbres Observatory's 2-m telescope on Haleakal\={a}, Hawai'i. They were reduced with \texttt{floyds\_pipeline},\footnote{\href{https://github.com/lcogt/floyds\_pipeline}{https://github.com/lcogt/floyds\_pipeline}} a PyRAF-based pipeline, using standard longslit spectral reduction methods as described by \citet{Valenti14}. These spectra are shown in Figure~\ref{fig:spec} and are available through the Weizmann Interactive Supernova Data Repository (WISeREP;\footnote{\href{https://www.wiserep.org/}{https://www.wiserep.org/}} \citealt{2012PASP..124..668Y}).


\section{Analysis}
\label{sec:analysis}

With only one epoch of early-phase NIR observations of SN 2021qvv, we must find a suitable literature 1991bg-like SN with extensively sampled early-phase NIR light curves to connect to our late-time \textit{HST} observations. Below, we show that the spectra (Section~\ref{subsec:classify}) and light curves (Section~\ref{subsec:params}) of SN 2021qvv are most similar to those of SN 2006mr, a well-studied 1991bg-like SN. A comparison of this object's NIR light curves with our late-time \textit{HST} data reveals no evidence of a NIR plateau out to 250 d (Section~\ref{subsec:plateau}).  

\subsection{Spectroscopic classification}
\label{subsec:classify}

Based on a spectrum from 2021 June 23, SN 2021qvv was first classified as a young SN Ia by the Young Supernova Experiment \citep{2021ApJ...908..143J,2021TNSCR2190....1A}. On 2021 July 27, the Zwicky Transient Facility (ZTF; \citealt{2019PASP..131a8002B}) classified the SN as a 1991bg-like object \citep{2021TNSCR2580....1C}. 

In Figure~\ref{fig:spec}, the three spectra of SN 2021qvv are compared to three well-studied SNe at similar phases, namely the 1991bg-like SNe 2005ke \citep{Silverman2012SNDB,2013ApJ...773...53F} and 2006mr \citep{2010AJ....140.2036S}, as well as the normal SN 2011fe \citep{2012ApJ...752L..26P,2013A&A...554A..27P,2020MNRAS.492.4325S}. The spectra of the comparison SNe have been normalized to match the Si \textsc{ii} absorption feature at 635.5 nm.

While the spectra of SN 2021qvv are clearly dissimilar to those of the normal SN 2011fe, they are nearly identical to those of SNe 2005ke and 2006mr. The conspicuous 400--440~nm Ti~\textsc{ii} band that sets 1991bg-like spectra apart from those of normal SNe Ia \citep{1992AJ....104.1543F} is apparent in the spectra of both SNe starting immediately after peak. This comparison confirms the ZTF classification of SN 2021qvv as a 1991bg-like SN Ia. 

Of the two 1991bg-like SNe, SN 2021qvv more closely resembles SN 2006mr, even though SN 2021qvv is brighter than SN 2006mr at wavelengths bluewards of $\approx 550$ nm. We attribute this difference to host-galaxy reddening that afflicts SN 2006mr (Section~\ref{subsubsec:red}). There are some notable differences between SN 2021qvv and SN 2005ke, such as the deeper Ti~\textsc{ii} trough in its pre-maximum spectrum and several features within the Fe-group complex in the +5 d spectrum. For more on the similarities and differences between these SNe, see Sections~\ref{subsubsec:sbv}--\ref{subsubsec:absmag}, below. 

\subsection{Photometric classification}
\label{subsec:params}

We continue the exercise of classifying SN 2021qvv by studying its light curves and comparing them to those of other 1991bg-like SNe.

\begin{table}
 \caption{Properties of SN 2021qvv and its host galaxy, NGC 4442}\label{table:params}
 \begin{tabular}{lcc}
  \hline
  \hline
  Property & Value & Ref. \\
  \hline
  \multicolumn{3}{c}{SN 2021qvv} \\
  \hline
  $\alpha$ (J2000) & $12^\mathrm{h}28^\mathrm{m}02\fs98$ & (1) \\
  $\delta$ (J2000) & $+09\degr48\arcmin09\farcs9$ & (1) \\
  \multirow{2}{*}{$t(U)_\mathrm{max}$} & $59388.5 \pm 2.2$ d & GPR \\
                                       & $59393.1 \pm 0.6 \pm 0.3$ (sys) d & SNooPy \\
  $t(B)_\mathrm{max}$ & $59395.3 \pm 0.1$ d & GPR \\
  $t(V)_\mathrm{max}$ & $59397.4 \pm 0.1$ d & GPR \\
  \multirow{2}{*}{$t(g)_\mathrm{max}$} & $59396.3 \pm 0.2$ d & GPR \\
                                       & $59395.1 \pm 0.6 \pm 0.3$ (sys) d & SNooPy \\
  $t(r)_\mathrm{max}$ & $59398.6 \pm 0.1$ d & GPR \\
  $t(i)_\mathrm{max}$ & $59399.7 \pm 0.1$ d & GPR \\
  $t(J)_\mathrm{max}$ & $59398.1 \pm 0.6 \pm 0.3$ (sys) d & SNooPy \\
  $t(H)_\mathrm{max}$ & $59399.1 \pm 0.6 \pm 0.3$ (sys) d & SNooPy \\
  \multirow{2}{*}{$m(U)_\mathrm{max}$} & $14.3 \pm 0.2$ mag & GPR \\
                                       & $14.42 \pm 0.15 \pm 0.03$ (sys) mag & SNooPy \\
  $m(B)_\mathrm{max}$ & $14.51 \pm 0.01$ mag & GPR \\
  $m(V)_\mathrm{max}$ & $13.86 \pm 0.01$ mag & GPR \\
  \multirow{2}{*}{$m(g)_\mathrm{max}$} & $14.31 \pm 0.02$ mag & GPR \\
                                       & $14.16 \pm 0.045 \pm 0.014$ (sys) mag & SNooPy \\
  $m(r)_\mathrm{max}$ & $13.78 \pm 0.01$ mag & GPR \\
  $m(i)_\mathrm{max}$ & $13.91 \pm 0.02$ mag & GPR \\
  $m(J)_\mathrm{max}$ & $13.22 \pm 0.30 \pm 0.04$ (sys) mag & SNooPy \\
  $m(H)_\mathrm{max}$ & $13.07 \pm 0.35 \pm 0.03$ (sys) mag & SNooPy \\
  $\Delta m_{15}(B)$ & $2.05 \pm 0.03$ mag & GPR \\
  \multirow{2}{*}{$s_\mathrm{BV}$} & $0.28 \pm 0.05$ & GPR \\
                                   & $0.27 \pm 0.02 \pm 0.03$ (sys) & SNooPy \\
  $E(B-V)_\mathrm{MW}$ & 0.019 mag & (2) \\
  $E(B-V)_\mathrm{Host}$ & $0.0 \pm 0.0$ mag & This work \\
  $M(B)_\mathrm{max}$ & $-16.42 \pm 0.07$ mag & GPR \\
  $M(^{56}\mathrm{Ni})$ & $0.04$--$0.06~M_\odot$ & UVOIR \\
  \hline
  \multicolumn{3}{c}{NGC 4442} \\
  \hline
  Type & Lenticular & (3) \\
  $\alpha$ (J2000) & $12^\mathrm{h}28^\mathrm{m}03\fs883$ & (4) \\
  $\delta$ (J2000) & $+09\degr48\arcmin13\farcs36$ & (4) \\
  $z_\mathrm{helio}$ & $0.00183 \pm 0.00002$ & (5) \\
  $v_\mathrm{helio}$ & $547 \pm 5~\rm{km~s^{-1}}$ & (5) \\
  $\mu$ & $30.85 \pm 0.07$ mag & This work \\
  $d$   & $14.8\pm 0.5$ Mpc & This work \\
  \hline
  \multicolumn{3}{l}{\textbf{Notes.} Peak magnitude uncertainties formally lower than 0.01 mag} \\
  \multicolumn{3}{l}{have been rounded up.} \\
  \multicolumn{3}{l}{\textbf{References:} (1) \citet{2021TNSTR2173....1T}; (2) \citet{2011ApJ...737..103S};} \\
  \multicolumn{3}{l}{(3) \citet{1991rc3..book.....D}; (4) \citet{2008AJ....135.1837R};} \\
  \multicolumn{3}{l}{(5) \citet{2011MNRAS.413..813C}.}
 \end{tabular}
\end{table}

\begin{figure}
 \begin{tabular}{c}
  \includegraphics[width=0.475\textwidth]{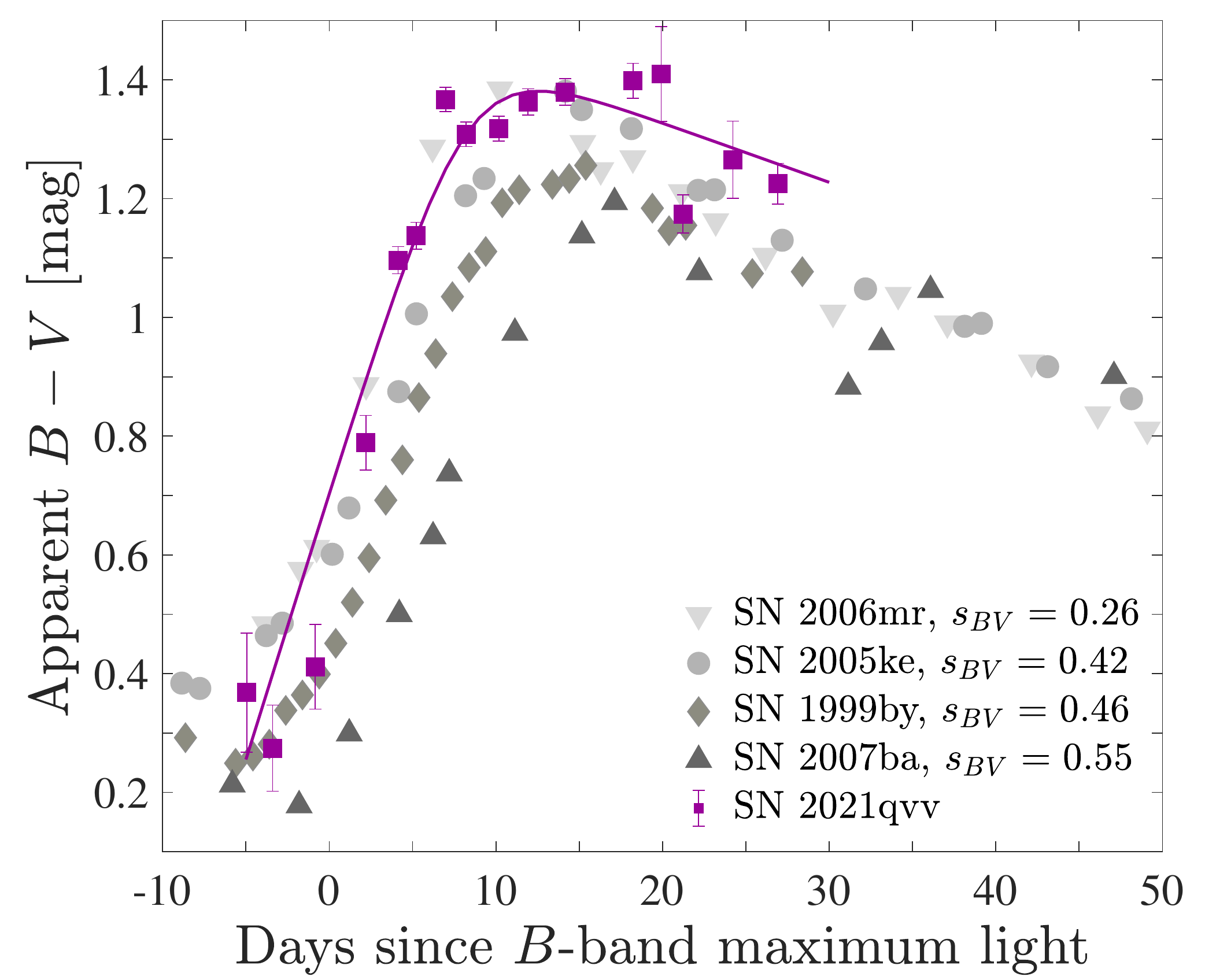} \\
  \includegraphics[width=0.475\textwidth]{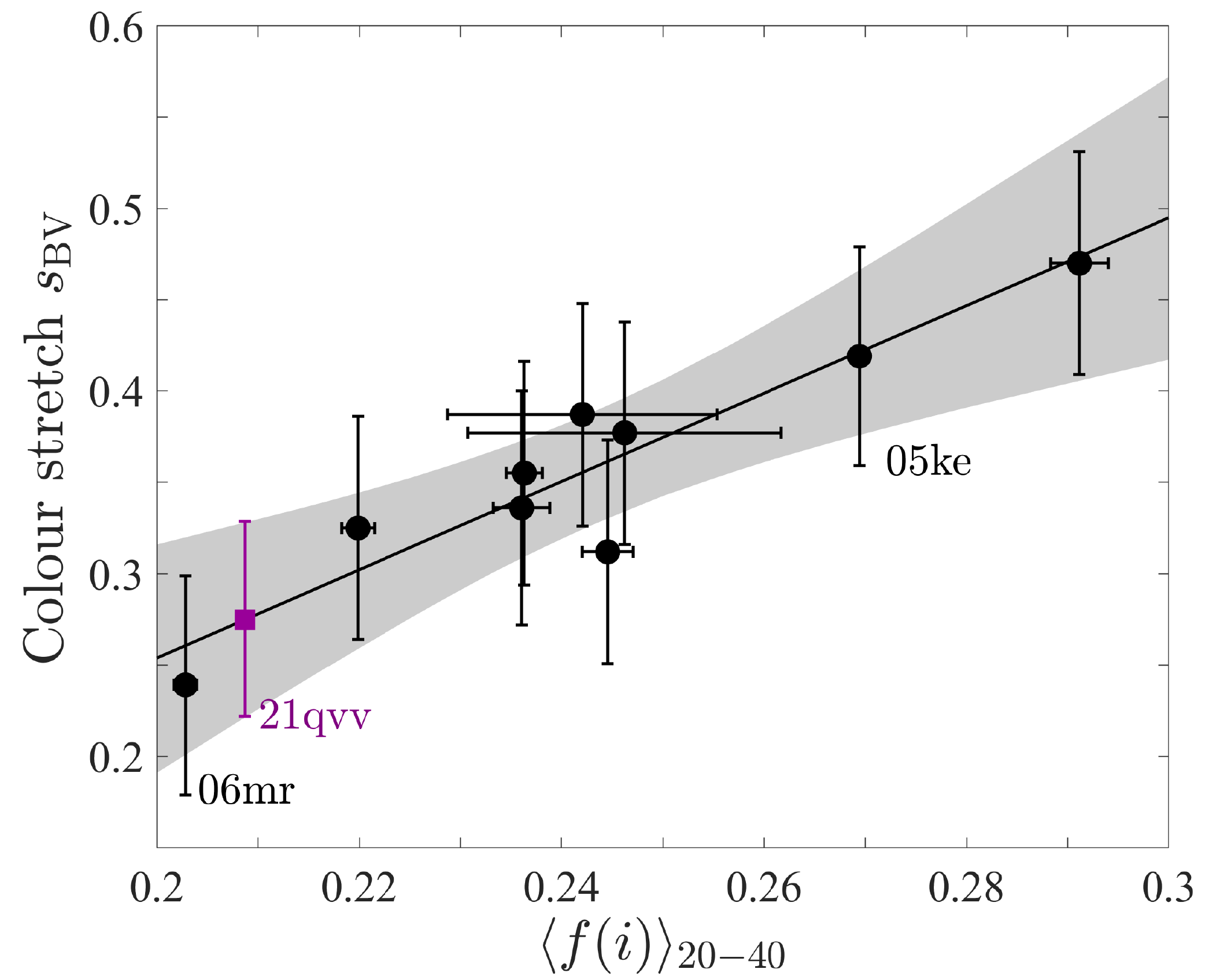} \\
 \end{tabular}
 \caption{Determining $s_\mathrm{BV}$. Top: The $B-V$ colour curve of SN 2021qvv (purple squares) is similar to that of SN 2005ke (circles). The data for SN 1999by were taken from \citet{2010ApJS..190..418G} while the data for SNe 2005ke, 2006mr, and 2007ba were taken from \citet{2017AJ....154..211K}. $s_{\mathrm BV}$ values for these SNe were taken from \citet{2014ApJ...789...32B} and \citet{2017AA...602A.118D}. Bottom: The correlation between $s_\mathrm{BV}$ and $\langle f_\lambda(i) \rangle_\mathrm{20-40}$ provides an $s_\mathrm{BV}$ value similar to that of SN 2006mr. The shaded area denotes the $1\sigma$ confidence region around the best-fitting first-order polynomial curve.}
 \label{fig:color}
\end{figure}

\subsubsection{Light-curve fits}

Standard light-curve fitters, such as MLCS2k2 and SNooPy \citep{2007ApJ...659..122J,2011AJ....141...19B}, classify SN 2021qvv as a 1991bg-like SN but disagree on the values of the various light-curve parameters, including the date and magnitude of $B$-band maximum light. MLCS2k2 is optimised to fit normal SNe Ia and does not contain many templates of 1991bg-like SNe. SNooPy is trained on a wider range of templates, including those of underluminous SNe Ia, but provides questionable fits to our $V$- and $r$-band light curves (Figure~\ref{fig:lc_LCO}).  

As a consequence, we fit our early, ground-based observations using Gaussian process regression (GPR). This is done using the Matlab R2023a \texttt{fitrgp} routine with the default parameters. The resultant fits provide the date and magnitude of maximum light in each filter. To estimate the uncertainties on these parameters, we repeat the fit in each filter 100 times, each time varying our photometry randomly according to the measurement uncertainties. The mean and standard deviation of the results are then taken as the estimated peak date and its $1\sigma$ uncertainty. The same method is applied to estimating the apparent peak magnitude in each filter. Magnitude uncertainties smaller than $0.01$ mag are rounded up. The GPR fits are shown in Figure~\ref{fig:lc_LCO}, and the resultant parameters are summarised in Table~\ref{table:params}.  

The `max model' used in the SNooPy fits provides a single date of maximum light ($59395.1 \pm 0.6 \pm 0.3$ (sys) d). As a result, the reported peak apparent magnitude in each band is the magnitude at the time of $B$-band maximum light, not the magnitude on the date in which the SN peaks in each filter (for details of the `max model,' see \citealt{2010AJ....140.2036S} and \citealt{2011AJ....141...19B}). This assumption does not hold for 1991bg-like SNe, in which the light curves in successively redder filters peak at progressively later times. Here, we extract the peak data and magnitude in each filter from the fits produced by SNooPy and assume that the formal uncertainties reported by SNooPy hold for these values as well. Because the SNooPy fits are overall similar to our GPR fits, we only report the peak date and apparent magnitude in $U$, $g$, $J$, and $H$, where the SNooPy fits offer additional information to that provided by the GPR fits. 

According to our GPR fits, SN 2021qvv reached $B$-band maximum light on MJD $59395.3 \pm 0.1$ d, with $m(B)_\mathrm{max} = 14.51 \pm 0.01$ mag. As seen in other 1991bg-like SNe, the date of maximum light arrives progressively later in redder filters. In our analysis below, we adopt these and other GPR-derived values. We caution not to place too much confidence in our $U$-, $g$-, $J$-, and $H$-band fits (from both the GPR and SNooPy fits), as the light curves do not provide good coverage of the peaks in those filters.

\subsubsection{Reddening}
\label{subsubsec:red}

The Galactic line-of-sight reddening towards SN 2021qvv is $E(B-V)_\mathrm{MW}=0.019$ mag \citep{2011ApJ...737..103S}. Based on its colors, location within a lenticular galaxy, and lack of narrow Na ID features (Figure~\ref{fig:spec}), we estimate no host-galaxy reddening. This is consistent with several other 1991bg-like SNe found in lenticular and elliptical galaxies (e.g., \citealt{2022ApJ...928..103H}).

\subsubsection{Colour stretch}
\label{subsubsec:sbv}

\citet{2014ApJ...789...32B} have shown that the shape of the $B-V$ colour curve of normal and underluminous SNe Ia is correlated with other light-curve parameters, such as $t(i)_\mathrm{max}$ or the strength of the second maximum in the $i$ band. They parameterised the time of $B-V$ maximum, divided by 30 d, as the `colour stretch' parameter $s_\mathrm{BV}$. The top panel of Figure~\ref{fig:color} shows the apparent $B-V$ colour curve of SN 2021qvv along with a selection of colour curves of 1991bg-like SNe Ia from the literature. The $B$- and $V$-band light curves were not corrected for Galactic or host-galaxy extinction, since \citet{2014ApJ...789...32B} have shown that the $B-V$ colour curves are insensitive to reddening, with extinctions up to $A_V=3$ mag resulting in changes of $<1$ per cent in $s_\mathrm{BV}$ (see their section 3.2).

We follow \citet{2014ApJ...789...32B} and fit the $B-V$ colour curve of SN 2021qvv with the model:
\begin{equation}
 y(t) = \frac{(s_{\mathrm 0}+s_{\mathrm 1})}{2}+\frac{(s_{\mathrm 1}-s_{\mathrm 0})\tau}{2}\mathrm{log}\left[\mathrm{cosh}\left(\frac{t-t_{\mathrm 1}}{\tau}\right)\right]+c+f_n(t,t_{\mathrm 0}),
\end{equation}
where $t$ is time; $s_{\mathrm 0}$ and $s_{\mathrm 1}$ are the slopes before and after the time at which the function curves, $t_{\mathrm 1}$, respectively; $\tau$ is the timescale over which the curving occurs; $c$ is a constant; and $f_n(t,t_{\mathrm 0})$ is a $n$-order polynomial for $t<t_{\mathrm 0}$ and 0 otherwise. We note that the original equation in \citet{2014ApJ...789...32B} contained two typos, which have been corrected here.

In the limit $\tau \to 0$ and $s_{\mathrm 0} \to s_{\mathrm 1}$, or just $s_{\mathrm 0} \to s_{\mathrm 1}$, the time at which the $B-V$ colour curve reaches maximum, $t_\mathrm{max}$, equals $t_{\mathrm 1}$. Otherwise,
\begin{equation}
 t_\mathrm{max} = t_{\mathrm 1} + \frac{\tau}{2}\mathrm{log}\left(\frac{-s_{\mathrm 0}}{s_{\mathrm 1}}\right).
\end{equation}
In either case, $s_\mathrm{BV} = t_\mathrm{max}/30$.

Given the scatter of our $B-V$ measurements, we dispense with the $f_n(t,t_{\mathrm 0})$ component of the fit and find best-fitting values of $s_0=0.09\pm0.01$, $s_1=-0.010\pm 0.003$, $\tau=3.9\pm1.3$ d, $t_1=8.2\pm1.2$ d, and $c=0.98\pm0.05$ mag. We note that, in order to converge, the fit requires the addition of an extra $\pm0.1$ mag to the uncertainties of our $B-V$ measurements to properly account for their scatter. These best-fit parameters result in $t_\mathrm{max} = 12.3\pm1.5$ d, which translates to $s_\mathrm{BV} = 0.41\pm0.05$.

A different way to estimate the colour stretch of SN 2021qvv is to study the correlation between $s_\mathrm{BV}$ and $\langle f_\lambda(i)\rangle_{20-40}$, which parameterises the strength of the second maximum in the $i$ band \citep{2001AJ....122.1616K,2019MNRAS.485.2343P}. While slower-evolving 1991bg-like SNe Ia exhibit a second peak in the $i$ band similar to that seen in normal SNe Ia, that bump gradually merges with the primary peak in SNe with faster-evolving light curves. 

We use GPR to fit the peaks and the 20--40 d range of the $i$-band light curves of 16 1991bg-like SNe Ia observed by the Carnegie Supernova Project \citep{2010AJ....139..519C,2011AJ....142..156S,2017AJ....154..211K}. Seven objects lack pre-maximum observations and are discarded, leaving us with nine SNe with $\langle f_\lambda(i)\rangle_\mathrm{20-40}$ values (namely, SNe 2005bl, 2005ke, 2006mr, 2007N, 2007ax, 2009F, 2006bd, 2007al, and 2008bt). The $s_\mathrm{BV}$ values of these SNe are taken from table 3 in \citet{2017AJ....154..211K}. We find $s_\mathrm{BV} = (2.4 \pm 0.8)\langle f_\lambda(i)\rangle_\mathrm{20-40} -(0.2 \pm 0.2)$ with $\chi^2/\mathrm{DOF}=1.3/7$. For SN 2021qvv, we measure $\langle f_\lambda(i)\rangle_\mathrm{20-40}=0.209 \pm 0.004$. As shown in the bottom panel of Figure~\ref{fig:color}, this correlation provides an $s_\mathrm{BV}$ value of $0.28 \pm 0.05$. This value is consistent with the value of $0.27 \pm 0.02 \pm 0.03$ (sys) derived by SNooPy.

An $s_\mathrm{BV}$ value of $0.41\pm0.05$ would put SN 2021qvv in the company of SN 2005ke ($s_\mathrm{BV}=0.419\pm0.003$), while a value of $0.28\pm0.05$ would make it more similar to SN 2006mr ($s_\mathrm{BV}=0.260\pm0.004$). Both SNe were fit by \citet{2014ApJ...789...32B}. 

\begin{figure*}
 \includegraphics[width=0.9\textwidth]{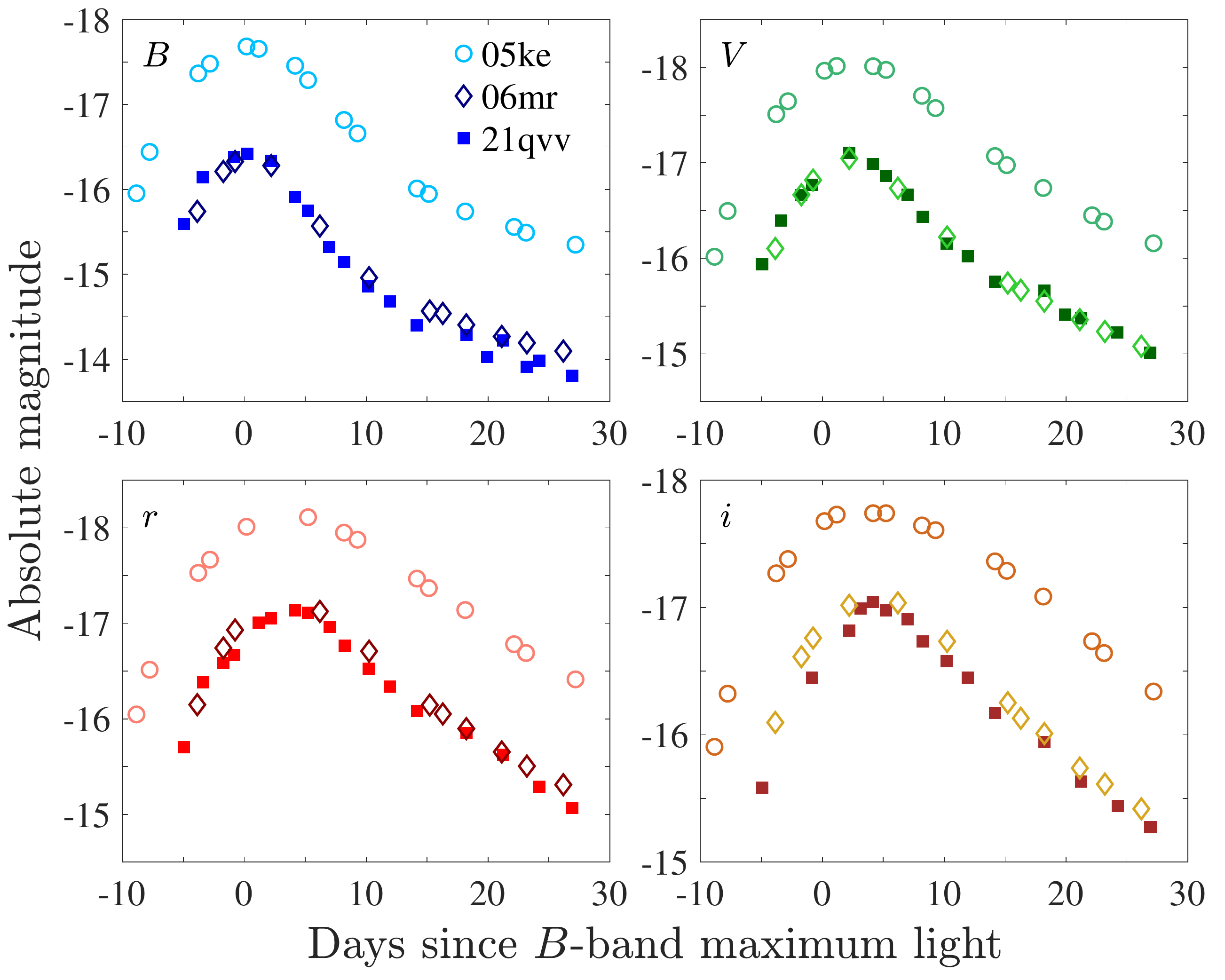}
 \caption{Comparison of the $B,~V,~r,$- and $i$-band light curves of SNe 2005ke (circles), 2006mr (diamonds), and 2021qvv (squares). Except in the $B$ band, the light curves of SN 2021qvv are more similar to those of SN 2006mr than those of SN 2005ke. The differences in magnitude between SNe 2006mr and 2021qvv are within the uncertainties of their peak $B$-band absolute magnitudes.}
 \label{fig:lc_comp}
\end{figure*}

A comparison of the $BVri$ light curves of SNe 2005ke, 2006mr, and 2021qvv reveals that SN 2021qvv evolves appreciably faster than SN 2005ke. Instead, as shown in Figure~\ref{fig:lc_comp}, SN 2021qvv is nearly a doppelg\"{a}nger of SN 2006mr. The only difference between these SNe lies in the $B$-band light curve, which then produces a wider $B-V$ colour curve than that of SN 2006mr. 

\begin{table*}
 \caption{Comparison between SNe 2005ke, 2006mr, and 2021qvv}\label{table:comp}
 \begin{tabular}{lccc}
  \hline
  \hline
  & 2005ke & 2006mr & 2021qvv \\
  \hline
  $m(B)_\mathrm{max}$ (mag)    & $14.90\pm0.01$  & $15.43\pm0.03$  & $14.51\pm0.01$ \\
  $\mu$ (mag)                  & $31.94\pm0.13$  & $31.36\pm0.04$  & $30.85\pm0.07$ \\
  $E(B-V)_\mathrm{MW}$ (mag)   & 0.020           & 0.018           & 0.019 \\
  $E(B-V)_\mathrm{Host}$ (mag) & $0.263\pm0.033$ & $0.089\pm0.039$ & $\cdots$ \\
  $R_V$                        & 1.0             & 2.9             & $\cdots$ \\
  $M(B)_\mathrm{max}$ (mag)    & $-17.68\pm0.13$ & $-16.36\pm0.12$ & $-16.42\pm0.07$ \\
  $s_\mathrm{BV}$              & $0.419\pm0.003$ & $0.260\pm0.004$ & $0.28\pm0.05$ \\
  \hline
 \end{tabular}
\end{table*}

\subsubsection{Peak absolute magnitude}
\label{subsubsec:absmag}

Another way to settle whether SN 2021qvv is more akin to SN 2005ke or to SN 2006mr is to compare their peak absolute $B$-band magnitudes. We use the same GPR technique described above to measure the peak apparent $B$-band magnitudes of SNe 2005ke and 2006mr from observations published by \citet{2017AJ....154..211K}. Table~\ref{table:comp} presents these values, along with distance moduli calculated below and Galactic \citep{2011ApJ...737..103S} and host-galaxy $E(B-V)$ and $R_V$ values \citep{2014ApJ...789...32B}.

Distance moduli of the host galaxy of SN 2021qvv, NGC 4442, have been measured using the Tully-Fisher \citep{2007AA...465...71T}, surface-brightness fluctuations \citep{2007ApJ...655..144M,2009ApJ...694..556B}, and globular-cluster radius and luminosity function techniques \citep{2005ApJ...634.1002J,2007ApJS..171..101J,2010ApJ...717..603V}. We calculate a weighted mean of the measurements reported by the works listed above and derive $\mu_\mathrm{21qvv} = 30.85\pm0.07$ mag.

In the same manner, we calculate weighted means of the distance moduli measured for the host galaxies of SNe 2005ke and 2006mr and find $\mu_{\mathrm{05ke}}=31.94 \pm 0.13$ mag (using values from \citealt{2001ApJ...546..681T,2002AA...393...57T,2013AJ....146...86T,2014MNRAS.444..527S,2016AJ....152...50T}) and $\mu_{\mathrm{06mr}} = 31.36\pm0.04$ mag (based on post-2010 values from \citealt{2013AA...552A.106C,2014ApJ...792..129N,2013ApJ...765...94S,2014MNRAS.444..527S,2016AJ....152...50T}). 

The host galaxies of SNe 2005ke and 2006mr also have distance moduli derived by fitting the light curves of the SNe themselves. However, as we have noted above, SN light-curve fitters are not optimized for 1991bg-like events. Thus, it is not surprising that these fits produce a wide range of values. Fits to SN 2005ke have produced distance moduli in the range 31.92--32.06 mag \citep{2008ApJ...689..377W,2010ApJ...721.1608B}, which are consistent with our weighted average, as well as $31.09 \pm 0.11$ mag \citep{2013ApJ...773...53F}, which is nearly a whole magnitude brighter. There is also a range of $\approx 1$ mag in the distance moduli derived from the light curves of SN 2006mr. \citet{2010AJ....140.2036S} used SNooPy to measure a distance modulus of $31.25 \pm 0.07$ mag from the light curves of three normal SNe Ia that exploded in the same galaxy (consistent with our weighted average), but $31.84 \pm 0.15$ mag from the light curves of SN 2006mr. \citet{2013ApJ...773...53F} measured an even fainter $32.58 \pm 0.34$ mag. Because of this wide range of values, we have excluded SN-based measurements from our weighted averages.

After applying the \citet{1989ApJ...345..245C} extinction law to the reddening towards each SN and subtracting the resultant $A_B$ values and distance moduli from the $B$-band apparent magnitudes measured here, we estimate peak absolute $B$-band magnitudes of $M(B)_\mathrm{05ke} = -17.68 \pm 0.13$ mag, $M(B)_\mathrm{06mr} = -16.37 \pm 0.15$ mag, and $M(B)_\mathrm{21qvv} = -16.42 \pm 0.07$ mag. The uncertainties of these measurements are the root-mean-squares of the uncertainties on the apparent magnitudes, distance moduli, host-galaxy reddening, and $R_V$ values. The luminosity we measure for SN 2021qvv, which is consistent with that of SN 2006mr, makes it one of the dimmest SNe Ia observed to date. SN 2005ke, on the other hand, is roughly three times as luminous.

In sum, we find that SN 2021qvv is most similar to SN 2006mr. The two SNe share a consistent peak absolute $B$-band brightness, fast-evolving light curves, and similar spectra. The only deviation appears in the $B$ band starting at $\approx 10$ d past maximum light, where SN 2021qvv seems to be fainter than SN 2006mr. This difference, which might simply be due to the scatter in our measurements at that phase, dictates the shape of the $B-V$ colour curve, which then appears to be more similar to that of the slower-evolving, more luminous SN 2005ke.

Our analysis has shown that, in order to place a given SN within the continuum of 1991bg-like SNe, it is not enough to measure its colour stretch, $s_\mathrm{BV}$. Instead, a complete analysis of its spectra, light curves, and absolute magnitude is required.

\begin{figure*}
 \includegraphics[width=0.9\textwidth]{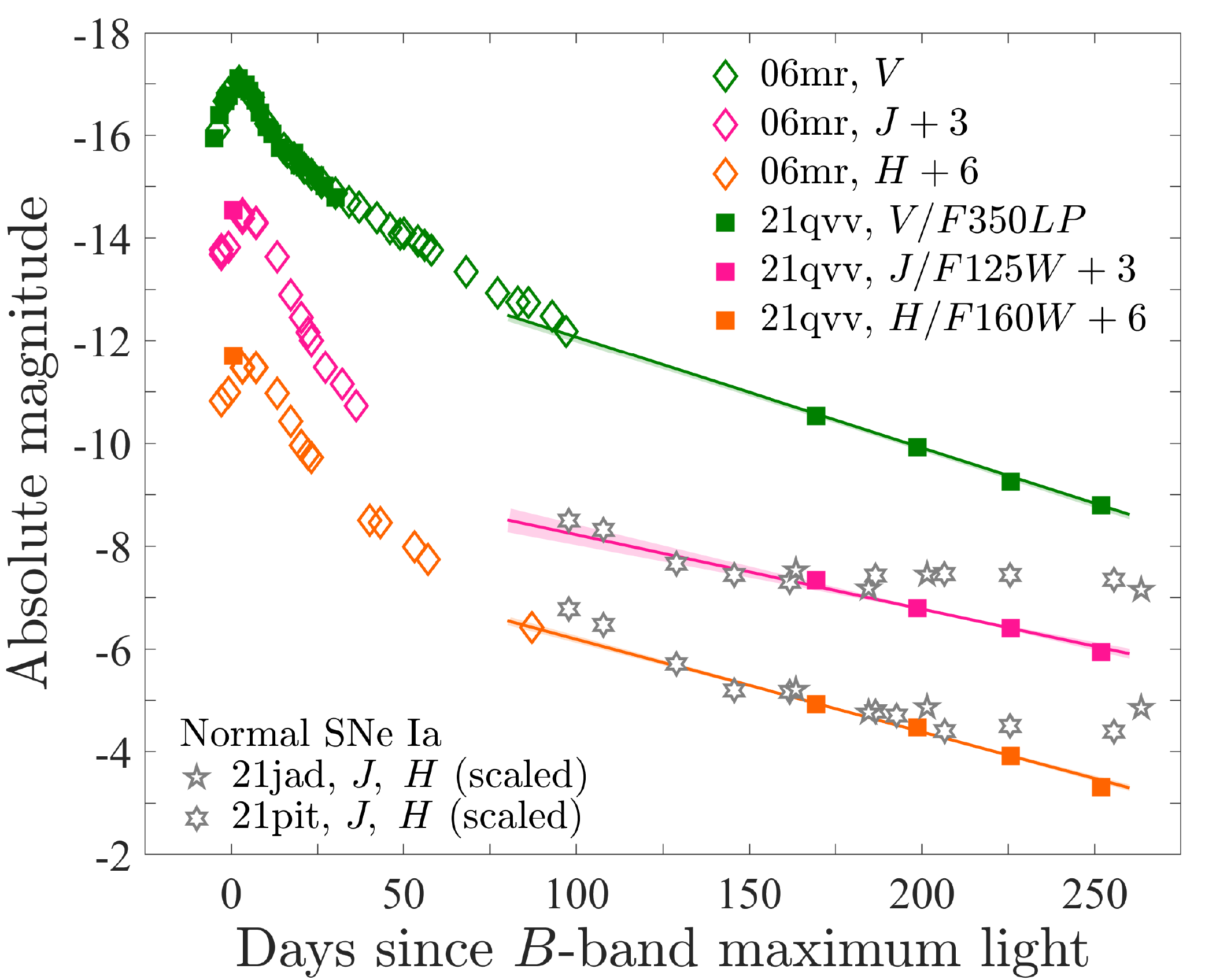}
 \caption{Comparison between the late-time \textit{HST} light curves of SN 2021qvv (squares), the early light curves of SN 2006mr (diamonds), and the NIR plateau phase of the normal SNe Ia 2021jad (pentagrams) and 2021pit (hexagrams). The solid curves are best-fitting polynomials to the late-time data, and the shaded regions around them represent the $2\sigma$ uncertainty regions of the fits. In the case of the \textit{F125W} light curve, the first measurement was excluded from the fit. Extrapolations of the fits show that, once we account for the similarity between SNe 2006mr and 2021qvv, there is no need for a late-time NIR plateau to connect between the early and late phases of the light curves. The late-time \textit{HST} observations of SN 2021qvv differ markedly from the observations of SNe 2021jad and 2021pit (scaled to match the light curves of SN 2021qvv at 169.5 d). Whereas the normal SNe Ia transition onto the NIR plateau, SN 2021qvv continues to decline in magnitude. Figure~\ref{fig:jad} presents a closer look at the late-time phase.}
 \label{fig:late}
\end{figure*}

\begin{figure}
 \includegraphics[width=0.475\textwidth]{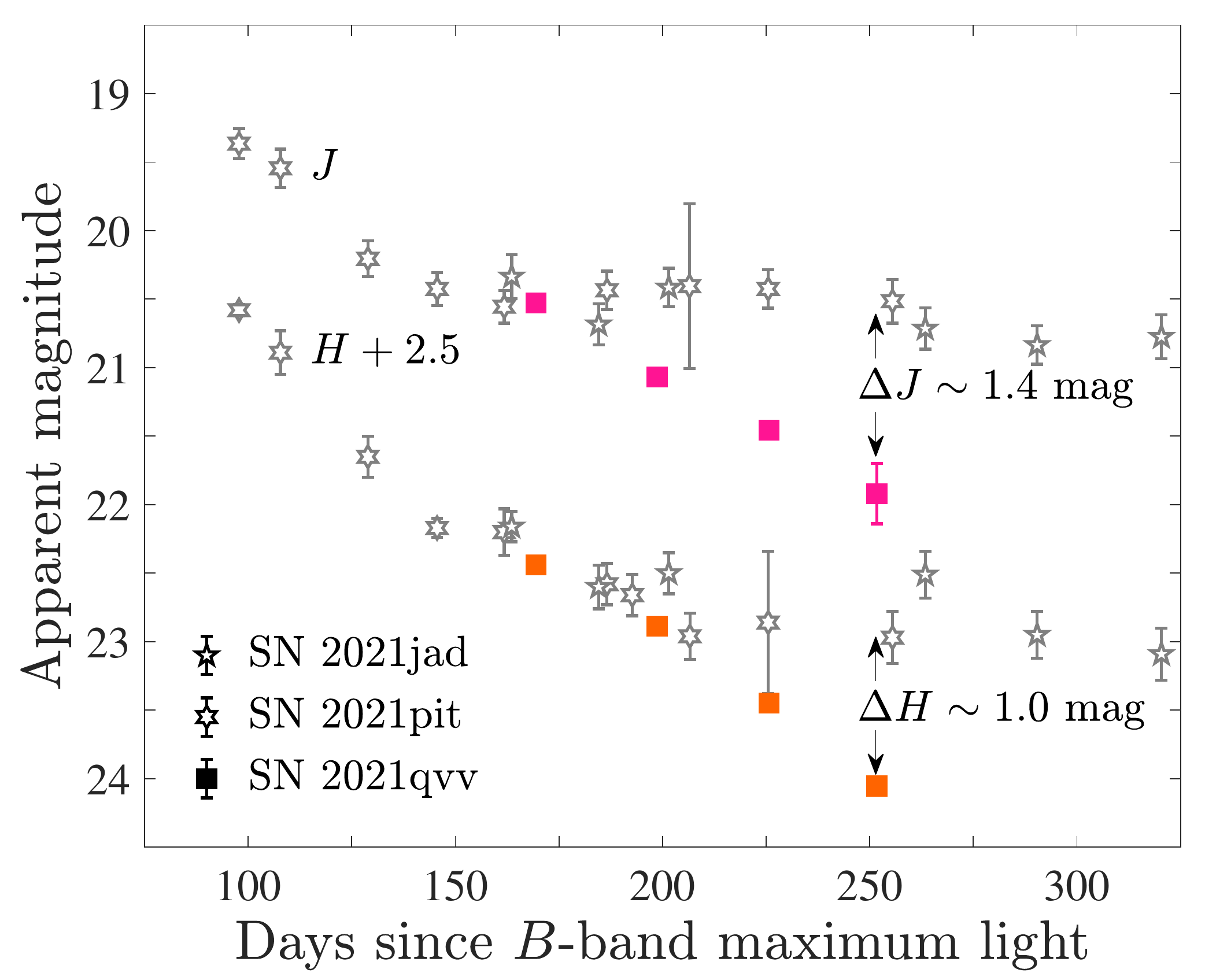}
 \caption{Comparison between the late-time NIR light curves of SN 2021qvv (squares) and normal SNe Ia 2021jad (pentagrams) and 2021pit (hexagrams). The light curves of SNe 2021jad and 2021pit have been shifted to match the light curves of SN 2021qvv at 169.5 d. While the normal SNe Ia transition onto the NIR plateau, SN 2021qvv continues to fade. By our final observation, SN 2021qvv is $\sim 1.4$ mag and $\sim 1$ mag fainter in \textit{F125W} and \textit{F160W}, respectively, than expected had it also plateaued.}
 \label{fig:jad}
\end{figure}

\subsection{No sign of a NIR plateau}
\label{subsec:plateau}

Now that we have established the similarity between SN 2021qvv and SN 2006mr, we can compare the late-time NIR observations of the former to the early NIR light curves of the latter. We begin by determining the late-time decline rate in each \textit{HST} band by fitting first-order polynomials to our measurements at 170--250 d. We find decline rates of $2.16 \pm 0.03$ mag/$100$ d in \textit{F350LP} ($\chi^2/\mathrm{DOF}=12/2$), $1.44 \pm 0.09$ mag/$100$ d in \textit{F125W} ($\chi^2/\mathrm{DOF}=0.16/2$), and $1.63 \pm 0.05$ mag/$100$ d in \textit{F160W} ($\chi^2/\mathrm{DOF}=27/2$). The high $\chi^2/\mathrm{DOF}$ values indicate that the formal uncertainties of our observations are underestimated. 

In Figure~\ref{fig:late}, we show the absolute $VJH$ light curves of SNe 2006mr and 2021qvv. We have not applied $S$ corrections, as we have no overlapping observations in similar ground-based and \textit{HST} filters, but these are expected to be $<0.5$ mag \citep{2023MNRAS.tmp..775D}. Our comparison of the \textit{F350LP} observations of SN 2021qvv to the $V$-band observations of SN 2006mr is based on the finding of \citet{2001ApJ...559.1019M} that, at $50<t<600$ d, a constant fraction of the luminosity of normal SNe Ia is emitted in the $V$ band, which dominates the 350--970 nm wavelength range spanned by the \textit{F350LP} band \citep{2016ApJ...819...31G}. We assume that the same is true of 1991bg-like SNe.

The early-phase $V$-band observations of SN 2021qvv are fully consistent with those of SN 2006mr. The $J$- and $H$-band observations at $0.6$ d, however, are $\sim 0.5$ mag brighter than those of SN 2006mr at the same phase. This difference could be attributed either to contamination from host-galaxy light or to the lack of $S$ corrections (see above). However, as we show below, it does not affect our analysis of the late-time \textit{HST} data.

A comparison between the early-phase light curves and extrapolations of the fits to the late-time \textit{HST} observations show a smooth transition in the $V$ and $H$ bands. SN 2006mr is not sampled in the $J$ band beyond 36 d, so we cannot connect its observations in this band to our late-time \textit{F125W} observations of SN 2021qvv.

Normal SNe Ia exhibit plateaus in the $J$ (\textit{F125W}) and $H$ (\textit{F160W}) bands between $\sim 150$--$500$ d with decline rates of $0.4\pm0.1$ mag$/100$ d and $0.5\pm0.1$ mag$/100$ d, respectively \citep{2023MNRAS.tmp..775D}. In Figures~\ref{fig:late} and \ref{fig:jad}, we compare our \textit{HST} observations to ground-based NIR observations of the normal SNe 2021jad and 2021pit taken at similar phases \citep{2023MNRAS.tmp..775D}. The $JH$ light curves of these SNe have been shifted in magnitude to match the \textit{F125W} and \textit{F160W} light curves of SN 2021qvv at 169.5 d. Even as the light curves of SNe 2021jad and 2021pit transition onto the NIR plateau, SN 2021qvv continues to fade. At 251.7 d, SN 2021qvv is $\sim 3.6$ and $\sim 2.5$ times dimmer in \textit{F125W} and \textit{F160W}, respectively, than it would have been had it plateaued.  

No statistical test is required to see that the decline rates of SN 2021qvv in the NIR are markedly different from those of normal SNe Ia. We conclude that, at least in the range 170--250 d, SN 2021qvv exhibited no measurable plateau. Furthermore, based on the agreement between the extrapolations of the late-time observations of SN 2021qvv and the early observations of SN 2006mr, we propose that SN 2021qvv did not go through a NIR plateau prior to 170 d, either.

\subsection{UVOIR light curve and $^{56}$Ni mass}
\label{subsec:bol}

\begin{figure}
 \includegraphics[width=0.475\textwidth]{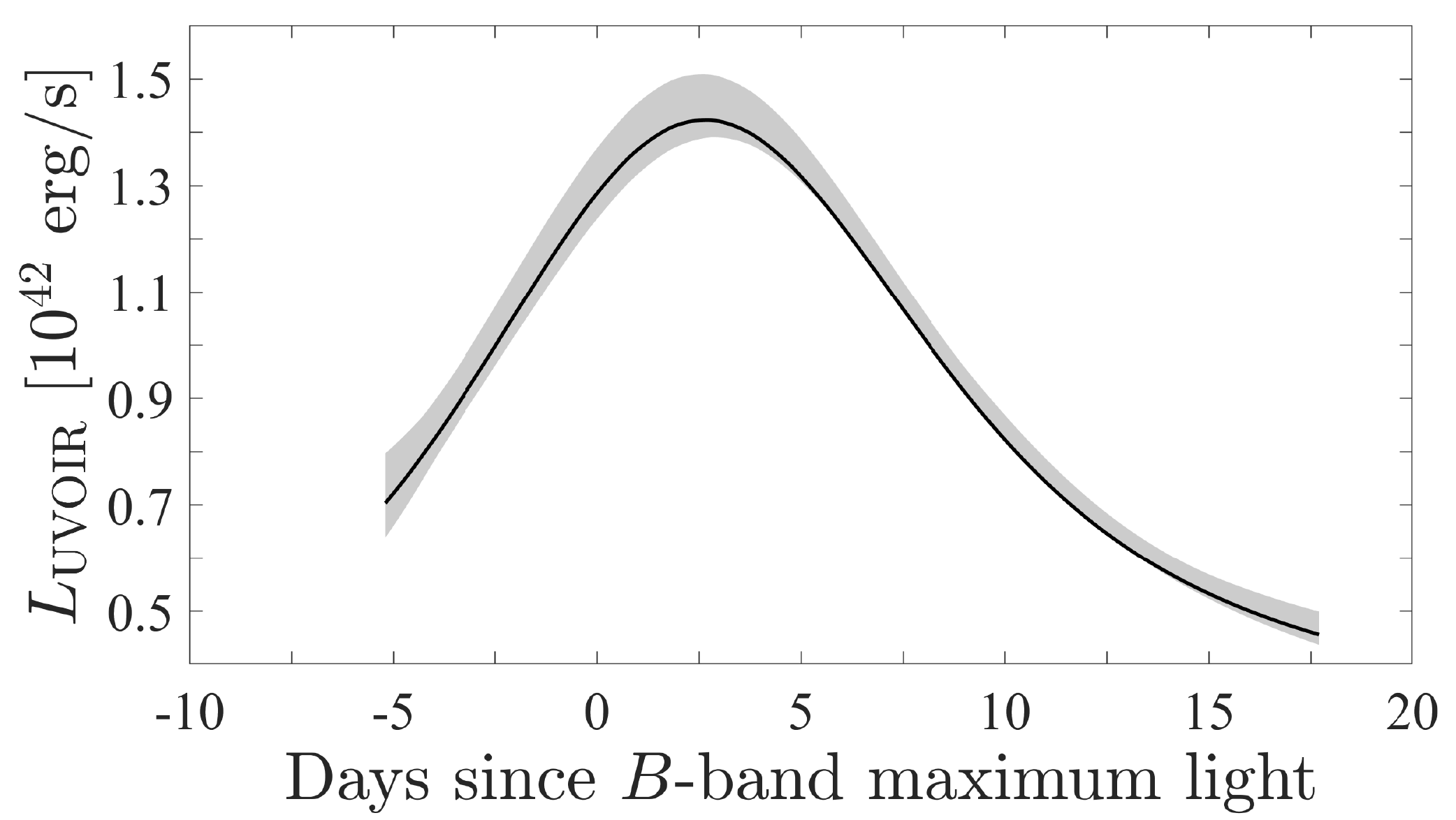}
 \caption{UVOIR luminosity curve of SN 2021qvv (solid curve) and $1\sigma$ uncertainty region (grey patch). The SN peaked at $L_\mathrm{max}=2.59^{+0.17}_{-0.15}~\mathrm{erg~s^{-1}}$, $\approx$ 2.7 d after $B$-band maximum light. The SNooPy fits to the NIR data have no uncertainties, so the formal uncertainty regions shown here are only lower estimates.}
 \label{fig:bol}
\end{figure}

Our $UBVgriJK$ observations of SN 2021qvv make it possible to construct the a UVOIR (or `pseudo-bolometric') light curve which, in turn, will yield an estimate of the mass of $^{56}$Ni produced by the explosion, $M(^{56}\mathrm{Ni})$. Since the SN was not observed with the full range of filters at each phase, we use our GPR fits to the $UBVgri$ filters and the SNooPy fits to the $J$- and $K$-band measurements presented in Figure~\ref{fig:lc_LCO}. At each phase, we convert our photometry to flux and construct an artificial, low-resolution spectrum that spans the wavelength range of our filters, 3100--18200 \AA. The artificial spectrum is then integrated and the result multiplied by $4\pi d^2$, where $d$ is the distance to the SN as derived from the distance modulus: $14.8\pm0.5$ Mpc. The resultant UVOIR light curve is shown in Fig.~\ref{fig:bol}.

We note that, while the UV and NIR light curves are not well sampled, each of these bandpasses provides only $\sim 10$ per cent of the flux that goes into the final UVOIR light curve, as seen in other 1991bg-like SNe \citep{2019MNRAS.483..628S}. We further note that, because the SNooPy-derived fits to the NIR bands lack uncertainties, the total uncertainty of the UVOIR light curve is underestimated.

To estimate $M(^{56}\mathrm{Ni})$, we follow Arnett's Rule \citep{1982ApJ...253..785A,1985Natur.314..337A,2000ApJ...530..744P,2000ApJ...530..757P}, according to which the maximum luminosity of an SN Ia, $L_\mathrm{max}$, is proportional to the instantaneous energy deposition rate from the radioactive decays within the expanding ejecta, $E_\mathrm{Ni}(t_R)$:
\begin{equation}
 L_\mathrm{max} = \alpha E_\mathrm{Ni}(t_R),
\end{equation}
where $t_R$ is the rise time to maximum light and the value of the proportional parameter, $\alpha$, is $\approx 1$; following \citet{2005A&A...431..423S}, we take $\alpha=1$. The energy produced by 1 $M_\odot$ of $^{56}$Ni is given by the equation \citep{2005A&A...431..423S}:
\begin{equation}
 E_\mathrm{Ni}(1~M_\odot) = 6.45\times10^{43}\mathrm{e}^{-t_R/8.8} + 1.45\times10^{43}\mathrm{e}^{-t_R/111.3},
\end{equation}
where $t_R$ is the rise time to maximum light and the denominators in the exponentials are the e-folding decay times for $^{56}$Ni and $^{56}$Co, respectively. The $^{56}$Ni mass of a given SN is then:
\begin{equation}
 M(^{56}\mathrm{Ni}) = \frac{L_\mathrm{max}}{E_\mathrm{Ni}(1~M_\odot)}~M_\odot.
\end{equation}

For normal SNe Ia, the mean rise time is $19\pm3$ d \citep{2005A&A...431..423S}, but underluminous SNe display faster rise times. \citet{2008MNRAS.385...75T}, for example, estimated a rise time of $14\pm1$ d for the underluminous SN 1999by. The UVOIR light curve derived here peaks $\approx 2.7$ d after $B$-band maximum light, which is $\approx 10$ d after the discovery of SN 2021qvv. This provides a rough lower limit on the rise time of the SN.

With a maximum UVOIR luminosity of $\approx 1.4\times10^{42}~\mathrm{erg~s^{-1}}$ (the underestimated formal uncertainties are $^{+0.09}_{-0.03}~\mathrm{erg~s^{-1}}$) and rise times of 10--15 d, we find that the $^{56}$Ni mass produced by SN 2021qvv lies in the range $0.04$--$0.06~M_\odot$. Using the same method, but a rise time of 19 d, \citet{2006A&A...460..793S} estimated $^{56}$Ni masses of $0.05$ and $0.09~M_\odot$ for SNe 1999by and 1991bg, respectively. Assuming a rise time of 14 d converts these values to $\sim 0.04$ and $\sim 0.07~M_\odot$, respectively. Hence, the $^{56}$Ni mass produced by SN 2021qvv is consistent with estimates derived from similar, though more luminous, 1991bg-like SNe.

\begin{figure*}
 \includegraphics[width=0.9\textwidth]{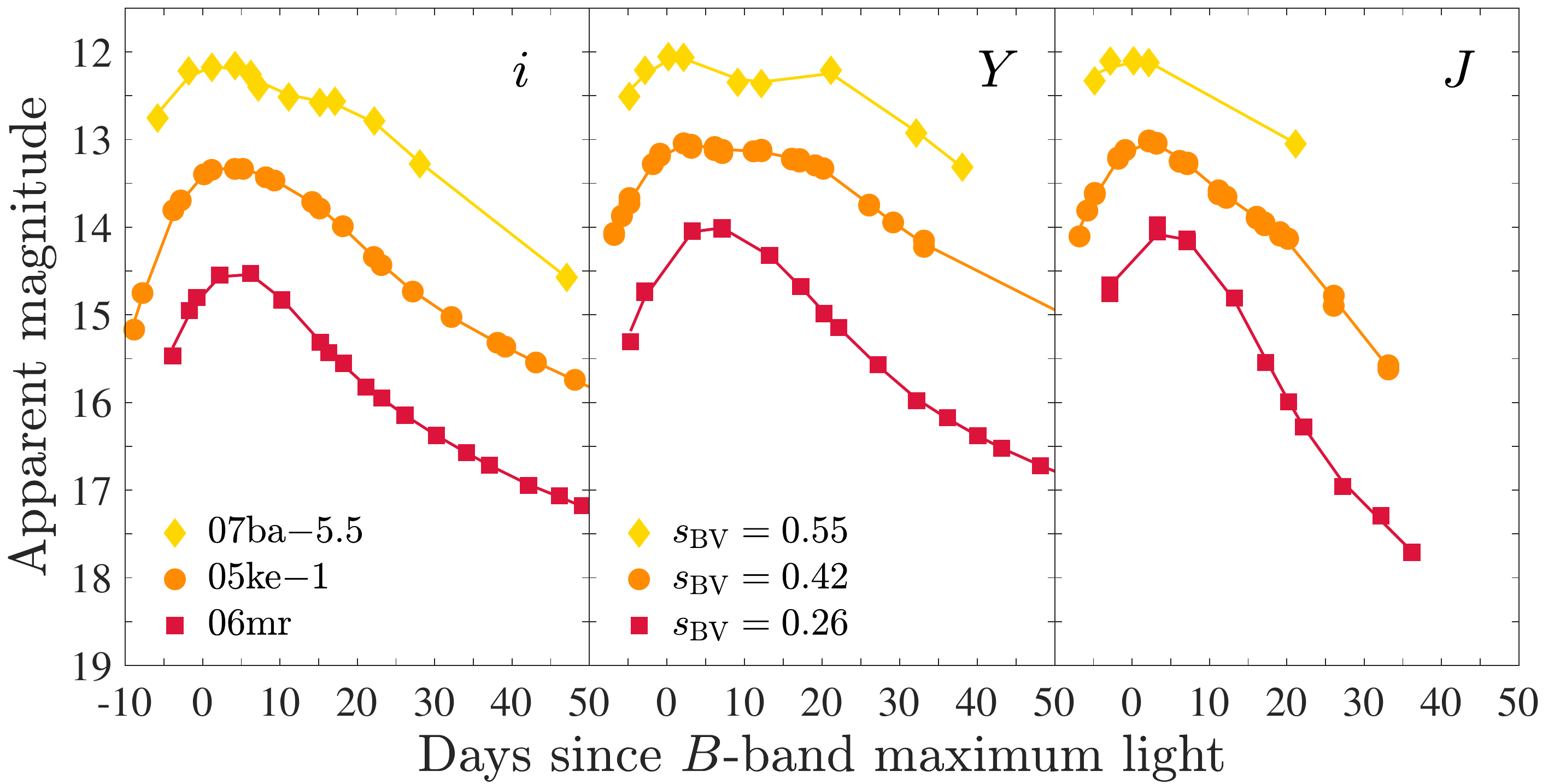}
 \caption{1991bg-like SNe with high $s_\mathrm{BV}$ values, such as SN 2007ba, exhibit second maxima in the $i$ and NIR bands similar to those seen in normal SNe Ia. As we progress to faster-evolving 1991bg-like SNe with lower $s_\mathrm{BV}$ values, such as SN 2005ke, the second maxima begin to blend into the initial peak. Finally, the second maxima disappear completely in the fastest-evolving SNe, such as SN 2006mr. This evolution may hint that, although we find no evidence for a late-time NIR plateau in SN 2021qvv, such a plateau may still exist in slower-evolving 1991bg-like SNe. }
 \label{fig:JHK}
\end{figure*}

\section{Discussion}
\label{sec:discuss}

This work is motivated by the preponderance of SN Ia subtypes and the need to associate each subtype with a progenitor and explosion scenario. It is widely agreed that the direct progenitor of all SNe Ia is a carbon-oxygen white dwarf (WD). As with all stars, WDs are found in different types of systems: alone; coupled with a second WD or a non-degenerate type of star; and in higher-order systems. Theory suggests various methods by which WDs in different stellar systems could be brought to explode (for the most recent review of the various progenitor channels, see \citealt{2023RAA....23h2001L}). 

The observed diversity of WD systems makes it hard to countenance that all SN Ia subtypes would arise from a single combination of progenitor and explosion scenario. At the same time, there is no reason to assume that each and every SN Ia subtype requires its own boutique set of physics. Instead, it is plausible that some SN Ia subtypes exist along a continuum, where variations in the composition of the exploding WD, the parameters of its stellar system, or the parameters of the explosion method (or some combination of all three) give rise to SNe Ia with a continuous range of spectroscopic and photometric properties.

There are reasons to assume that 1991bg-like SNe are distinct from normal SNe Ia. These include the unique spectroscopic features of these SNe, such as their telltale Ti~\textsc{ii} trough (Figure~\ref{fig:spec}); their apparent non-adherence to the peak width-luminosity relation (\citealt{Phillips1999,2004ApJ...613.1120G,2008MNRAS.385...75T}; but see \citealt{2014ApJ...789...32B}); and their preference for exploding in massive, passive galaxies with old stellar populations \citep{2001ApJ...554L.193H,2009ApJ...707.1449N,2017ApJ...837..121G,2019MNRAS.490..718B,2023arXiv230601088L}.

At the same time, there are also suggestions that 1991bg-like SNe Ia are but dimmer, faster-evolving siblings of normal SNe Ia. \citet{1995ApJ...455L.147N} and \citet{2017ApJ...846...15H} showed that varying the luminosity and temperature of a single ejecta structure could produce the spectroscopic properties of both normal and underluminous SNe Ia. \citet{2013MNRAS.429.2127B} showed that by varying the amount of $^{56}$Ni synthesized in the delayed detonation of a Chandrasekhar-mass WD, they could recreate the photometric and spectroscopic properties of both subtypes. In a similar vein, \citet{2018ApJ...854...52S,2021ApJ...909L..18S} showed that underluminous 1991bg-like SNe, normal SNe Ia, and overluminous 1991T-like SNe all arose from double detonations of WDs with varying initial masses between 0.85 and 1.1 $M_\odot$. This result was echoed by \citet{2019ApJ...873...84P,2021ApJ...906...65P}, who showed that double detonations of WDs with thin He shells ($0.01~M_\odot$) recreated many of the features seen in near-peak and nebular spectra of underluminous, normal, and overluminous SNe Ia.

On the face of it, the absence of a NIR plateau in SN 2021qvv should further set 1991bg-like SNe apart from normal SNe Ia. There is, however, an argument to be made that this is another sign that 1991bg-like SNe Ia exist on a continuum with normal SNe Ia.

Nebular NIR spectra of normal SNe show that, during the plateau phase, [Fe \textsc{iii}] emission lines fade away while [Fe \textsc{ii}] lines remain at a constant luminosity \citep{2018A&A...619A.102D,2018ApJ...861..119D,2018MNRAS.477.3567M}, suggesting that doubly-ionized iron ions are recombining to singly-ionized ions \citep{2023MNRAS.tmp..775D}. In this respect, the NIR plateau is reminiscent of the secondary NIR maximum seen in SNe Ia, which is thought to be the result of a `recombination wave' that sweeps through the ejecta. Starting out in the cooler outer layers of the ejecta, this wave recedes inwards. When the iron-rich inner regions of the ejecta begin to recombine, the iron-rich gas becomes fluorescent: high-energy photons scatter off the dense forest of UV iron lines and are redistributed to the NIR \citep{2000ApJ...530..757P,2006ApJ...649..939K,2015ApJ...814L...2F}. A similar effect is thought to be the reason for the existence of the late-time NIR plateau \citep{2004A&A...428..555S,2015ApJ...814L...2F,2020NatAs...4..188G}. 

The recombination of [Fe \textsc{iii}] to [Fe \textsc{ii}] provides two possible explanations for the absence of a NIR plateau in SN 2021qvv. First, one-dimensional, non-local thermodynamic equilibrium time-dependent radiative transfer simulations of four SN Ia ejecta models conducted by \citet{2018MNRAS.474.3187W} showed a marked difference in the late-time NIR spectra produced by the central detonation of a sub-Chandrasekhar-mass WD. Artificial spectra produced $\sim 100$ and $\sim 200$ d after bolometric maximum light showed doubly-ionized Fe and Co lines but no singly-ionized lines from these elements. The absence of these lines was explained by a larger fraction of local radioactive heating by positrons, which inhibits recombination. Without the recombination of [Fe \textsc{iii}] to [Fe \textsc{ii}], a SN produced by such an explosion scenario would not experience a late-time NIR plateau. This explanation, if borne out by spectroscopic observations of underluminous SNe Ia at late times, would strengthen the proposition that these SNe are the result of detonations of sub-Chandrasekhar-mass WDs.

A second explanation may be found in the observation that the NIR light curves of 1991bg-like SNe display a marked evolution from slower- to faster-evolving objects (Figure~\ref{fig:JHK}). The slowest-evolving 1991bg-like SNe, such as SN 2007ba ($s_\mathrm{BV}=0.55$), display secondary peaks in the $iYJ$ bands similar to those seen in normal SNe Ia. As we progress to faster-evolving 1991bg-like SNe, such as SN 2005ke ($s_\mathrm{BV}=0.42$), the secondary maxima shift to earlier times and begin to blend with the primary maxima. These features are completely absent from the fastest-evolving 1991bg-like SNe, such as SN 2006mr ($s_\mathrm{BV}=0.26$).

The evolution of the secondary NIR maximum may be related to the temperature of the SN ejecta, which are thought to be colder in 1991bg-like SNe than in normal SNe Ia (e.g., \citealt{1997MNRAS.284..151M,2011PASP..123..765D,2017ApJ...846...15H}). For example, the progressive narrowing of line widths seen in optical nebular spectra from normal to 1991bg-like SNe is interpreted as evidence that faster-evolving SNe Ia (with lower $s_\mathrm{BV}$ values) have progressively colder ejecta \citep{2022MNRAS.511.3682G,2023ApJ...945...27K}. In this scenario, as we transition from hotter (high $s_\mathrm{BV}$) to colder (low $s_\mathrm{BV}$) SNe, the iron recombination wave responsible for the second maximum occurs at earlier times, eventually merging with the primary maximum in the coldest 1991bg-like SNe \citep{2017hsn..book..317T}. 

In the same vein, perhaps below a certain temperature the ejecta of 1991bg-like SNe become too cold to produce the fluorescence mechanism that gives rise to the late-time NIR plateau. Spectroscopic modeling of 1991bg-like SNe---and especially of SNe 2006mr and 2021qvv---could reveal whether this is the case. Such modeling could also predict whether all 1991bg-like SNe are too cold for the fluorescence mechanism or whether there is a temperature threshold above which flux can once more be redistributed from the UV to the NIR. Future NIR observations of 1991bg-like SNe at 150--500 d would then reveal which is the case. Either the NIR plateau will never be observed in underluminous SNe or it will appear only in SNe with a certain $s_\mathrm{BV}$ value and above.

The results of such a test would also impact our understanding of normal SNe Ia. A detection of the NIR plateau in hotter 1991bg-like SNe would strengthen the suggestion that these underluminous SNe are an extension of normal SNe Ia. This would be doubly true if the appearance of the NIR plateau becomes more similar to that of normal SNe Ia as we progress from colder to hotter 1991bg-like SNe. Any progenitor and explosion scenario will then be required to recreate the observed properties of both normal and 1991bg-like SNe, lending more credence to works that manage to accomplish this feat (e.g., \citealt{2013MNRAS.429.2127B} and \citealt{2021ApJ...909L..18S}) over works that treat 1991bg-like SNe as a distinct subtype (e.g., \citealt{2010Natur.463...61P,2011AA...528A.117P,2017AA...602A.118D}). 


\section{Conclusions}
\label{sec:conclude}

In this work, we have presented optical and NIR observations of SN 2021qvv at early times ($-5$ to 30 d past $B$-band maximum) and at late times (170--250 d). The spectroscopy and early light curves of SN 2021qvv show that it belongs to the underluminous, 1991bg-like subtype of SNe Ia. SN 2021qvv closely resembles SN 2006mr, making it one of the dimmest ($M(B)_\mathrm{max} = -16.42 \pm 0.07$ mag) and fastest-evolving ($s_\mathrm{BV} = 0.28 \pm 0.05$) 1991bg-like SNe to date.

A comparison of the early NIR light curves of SN 2006mr to our late-time \textit{HST} observations of SN 2021qvv reveals no indication of the late-time NIR plateau seen in normal SNe Ia, at least out to 250 d past maximum light. The absence of the plateau is consistent with certain sub-Chandrasekhar-mass detonation models that inhibit the recombination of [Fe \textsc{iii}] to [Fe \textsc{ii}] that drives the NIR plateau \citep{2018MNRAS.474.3187W}.

We further argue that this absence does not preclude the appearance of a late-time NIR plateau in other 1991bg-like SNe. The secondary NIR maxima routinely observed in normal SNe Ia are also observed in slowly-evolving 1991bg-like SNe but gradually occur earlier in faster-evolving SNe until finally merging with the primary maximum in the fastest-evolving underluminous SNe Ia. This is thought to be the result of lower temperatures in the ejecta of faster-evolving 1991bg-like SNe. Since SN 2021qvv is one of the dimmest, fastest-evolving SNe observed to date, the lack of a plateau may also be connected to its presumably colder ejecta. 

Nebular NIR spectra of 1991bg-like SNe Ia will reveal whether the absence of the NIR plateau is due to a high ionization state of the ejecta, as predicted by \citet{2018MNRAS.474.3187W}. In parallel, spectroscopic modeling of fast-evolving 1991bg-like SNe Ia, including SNe 2006mr and 2021qvv, could reveal whether the fluorescence mechanism that is thought to give rise to the NIR plateau depends on temperature, and whether below a certain ejecta temperature it would cease to operate altogether. In parallel, a larger sample of 1991bg-like SNe observed in the NIR out to 500 d should reveal whether the NIR plateau is absent in all such objects, or whether there is a correlation between the appearance of the plateau and the temperature of the SN, as encapsulated in its colour stretch parameter, $s_\mathrm{BV}$. The existence of such a correlation would strengthen the proposition that 1991bg-like SNe and normal SNe Ia exist along a single continuum and, hence, share the same progenitor and explosion physics.


\section*{Acknowledgments}
We thank Chris Burns for his assistance fitting the colour curve of SN 2021qvv and the anonymous referee for their comments and suggestions. M.D. and K.M. are funded by EU H2020 ERC grant no. 758638. L.G. acknowledges financial support from the Spanish Ministerio de Ciencia e Innovaci\'on (MCIN), the Agencia Estatal de Investigaci\'on (AEI) 10.13039/501100011033, and the European Social Fund (ESF) `Investing in your future' under the 2019 Ram\'on y Cajal program RYC2019-027683-I and the PID2020-115253GA-I00 HOSTFLOWS project, from Centro Superior de Investigaciones Cient\'ificas (CSIC) under the PIE project 20215AT016, and the program Unidad de Excelencia Mar\'ia de Maeztu CEX2020-001058-M. The Aarhus supernova group is funded in part by an Experiment grant (\# 28021) from the Villum FONDEN, and by project grants (\#8021-00170B, 10.46540/2032-00022B) from the Independent Research Fund Denmark (IRFD). K.J.S. was in part supported by NASA/ESA {\it Hubble Space Telescope programs} \#15871 and \#15918. This work is based on data obtained with the NASA/ESA \textit{Hubble Space Telescope} through program GO-16884, as well as data obtained with the Las Cumbres global telescope network. The Las Cumbres Observatory group is supported by NSF grants AST-1911151 and AST-1911225, as well as \textit{HST}-GO-16884. This research has also made use of NASA's Astrophysics Data System and the NASA/IPAC Extragalactic Database (NED), which is funded by the National Aeronautics and Space Administration and operated by the California Institute of Technology. Finally, this work has made use of the Open Supernova Catalog \citep{2017ApJ...835...64G}. 


\section*{Data Availability}
The raw \textit{HST} data can be found in MAST at doi:\href{http://dx.doi.org/10.17909/psf9-7d70}{10.17909/psf9-7d70}. The ground-based images are archived by the Las Cumbres Observatory and can be accessed by request. The reduced spectra will be made available in WISeREP at \href{https://www.wiserep.org/}{https://www.wiserep.org/}. The reduced data and derived measurements in this article will be shared on request to the corresponding author.


\end{document}